\documentclass[twocolumn,twoside,slac_two]{revtex4}
\usepackage{graphicx}
\usepackage{fancyhdr}
\begin{document}

\title{Anomalous grouping of some short BATSE GRBs }

\author{Z. Bagoly} \affiliation{E\"otv\"os University, Budapest}
\author{L. G. Bal\'azs} \affiliation{Konkoly Observatory, Budapest}
\author{P. Veres} \affiliation{E\"otv\"os University, Budapest, Bolyai Military University, Budapest}
\author{A. M\'esz\'aros} \affiliation{Charles University, Prague}
\author{I. Horv\'ath} \affiliation{Bolyai Military University, Budapest}

\begin{abstract}
The power spectra of 457 short BATSE bursts were analyzed, focusing on the $64
ms$ lightcurves' tails in the low energy bands. Using MC simulations, 22 GRBs
were identified with unusually high harmonic power above $0.03 Hz$. 

The sky distribution of these bursts shows an extraordinarily strong dipole
moment with a $99.994\%$ significance. 
\end{abstract}

\maketitle

\thispagestyle{fancy}

\section{Data pre-processing} 

The shape of the gamma-ray burst's lightcurves of the BATSE Gamma-Ray Burst
Catalog \citep{mee00} carry an immense amount of information.  Here we focus on
the possible SGR-like contamination in the BATSE data.  We used data from the
BATSE DISCSC $64 ms$ dataset keeping in mind that the distant SGR signals could
be very similar to the short GRBs: a bright short peak and a - probably weak,
noisy - extended periodic emission in the low energy bands.

During the processing we've focused on the harmonic content of the GRB lightcurve. 
Therefore during the analysis  

\begin{itemize}

\item the three background-subtracted lowest energy bands (below $320$ keV)
were joined to reduce the noise  (the high energy ($ >320$ keV) channel behaves
differently \citep{bag98}). We used a quadratic background fit for
each channel.

\item FFT was used to calculate the power spectrum with a window size of 1024
samples ($65.536 s$). This sampling determines both the lowest frequency
($f=0.03052 Hz$, $T=32.768s$) and the highest ($f=7.8125 Hz$, $T=0.128s$). 

\item the first $8.192 s$ data was excluded from the signal after the trigger
($T0$): this ensures that even the strongest short GRB lightcurve will decay
and it does not interfere with any possible signal in the tail. 

\item we generated as many data windows as possible. The starting point of the
1024-sample-sized sliding window was increased with 256 samples ($16.384 s$)
until the window's endpoint reaches the end of the data.

\item to reduce the power spectrums sensitivity to edge effects a Blackman
window was used: this data windowing prevents leakage from the nearby frequency
components \cite{press}, and makes the power spectrum more regular. The window
also reduces any spurious effects originating from the non-perfect background
subtraction.

\item each data block was whitened before the FFT. No data winzorization was
applied.

\end{itemize}

\section{Power spectrum analysis}

Most of the data probably contains a Poissonian noise only, therefore to
determine the significance of any probable signal a Monte-Carlo simulation was
needed.

For each 1024 sample block random signals were generated 500 times by randomly
shuffling the data: the maximum value in the power spectrum of each MC
signal was compared with the real signal's maximum value. We select only bursts
where the real signal's maximum power is bigger than {\em any} maximum in the
MC signals.  The candidate trigger also required to produce such
harmonic power signal at least for two different window starting point
(overlapping windows were not excluded).

These requirements select 22 GRBs with unusually high harmonic power above
$0.03052 Hz$. Fig. 1. shows the power spectra of these bursts with the
corresponding window start times.

\begin{figure*}[t]
          \centering
\includegraphics[height=.656\columnwidth, angle=270]{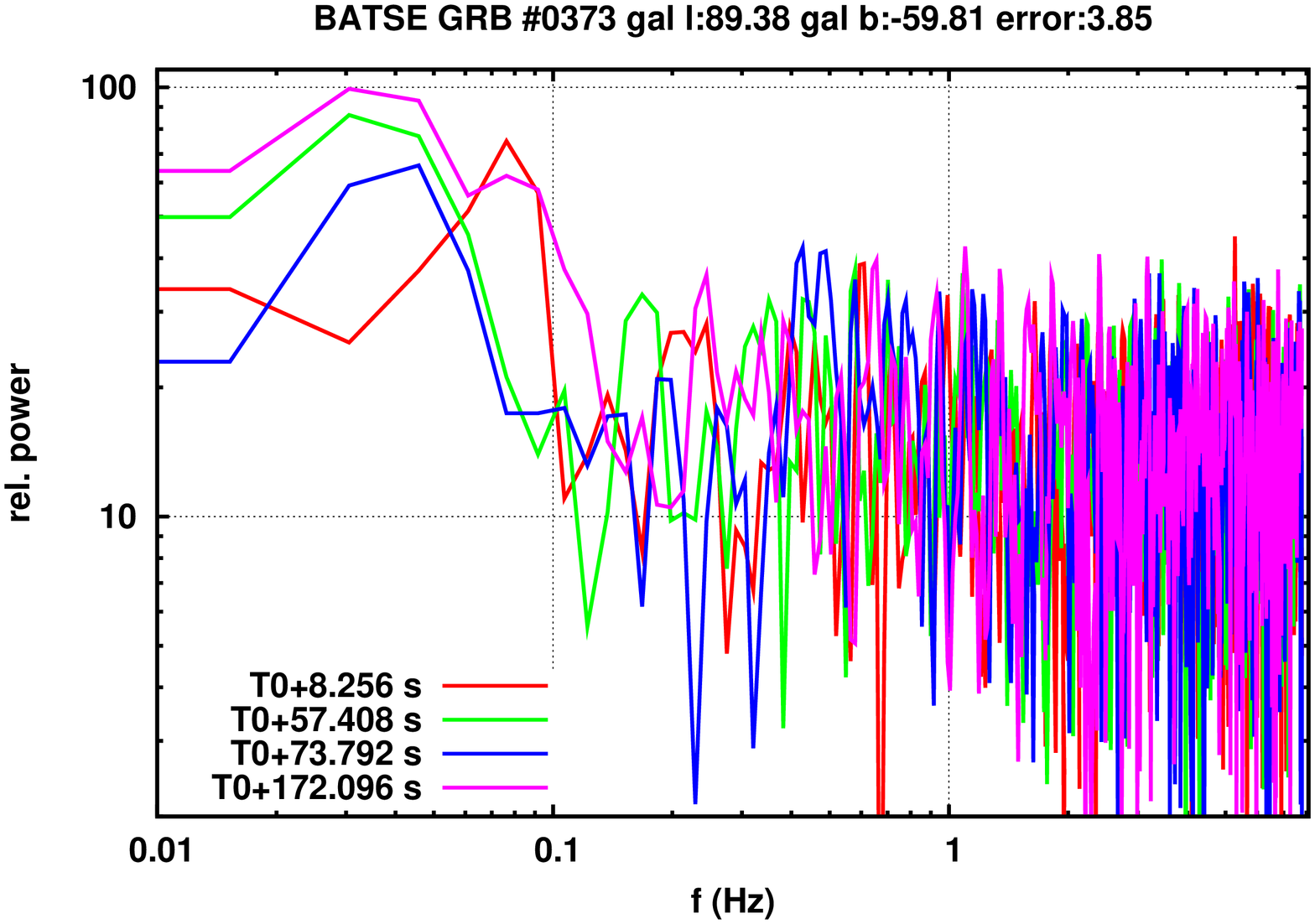}
\includegraphics[height=.656\columnwidth, angle=270]{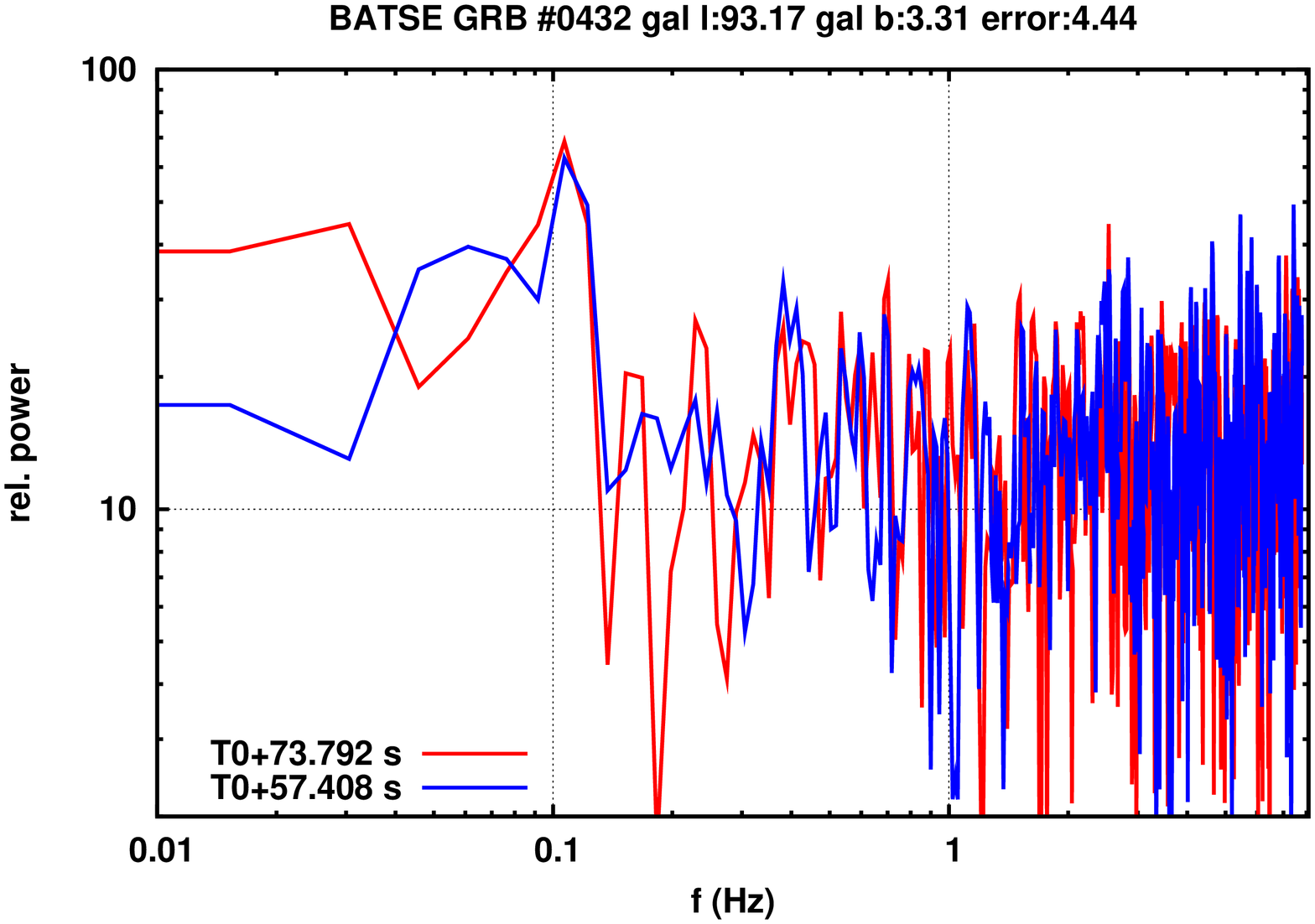}

\includegraphics[height=.656\columnwidth, angle=270]{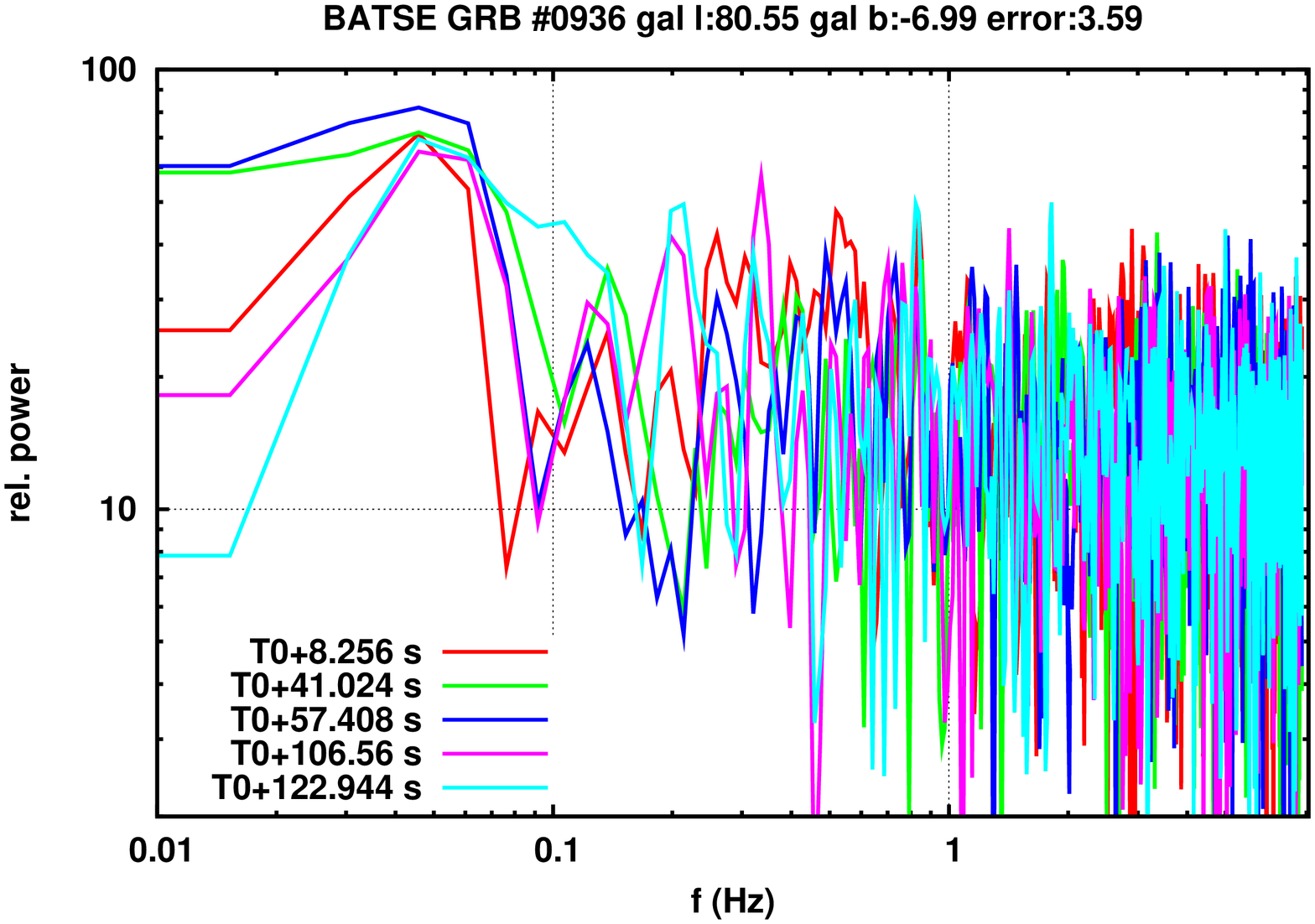}
\includegraphics[height=.656\columnwidth, angle=270]{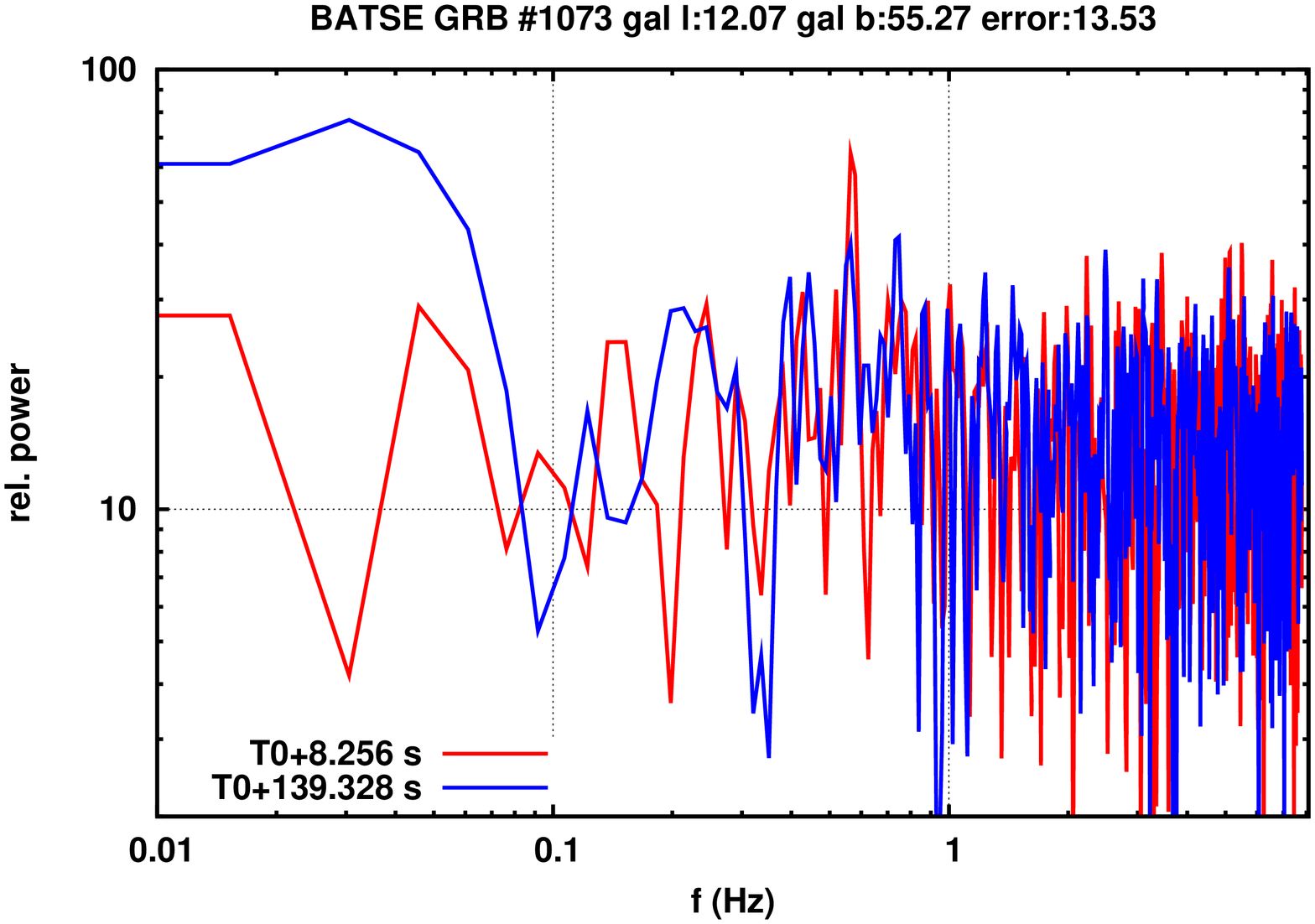}

\includegraphics[height=.656\columnwidth, angle=270]{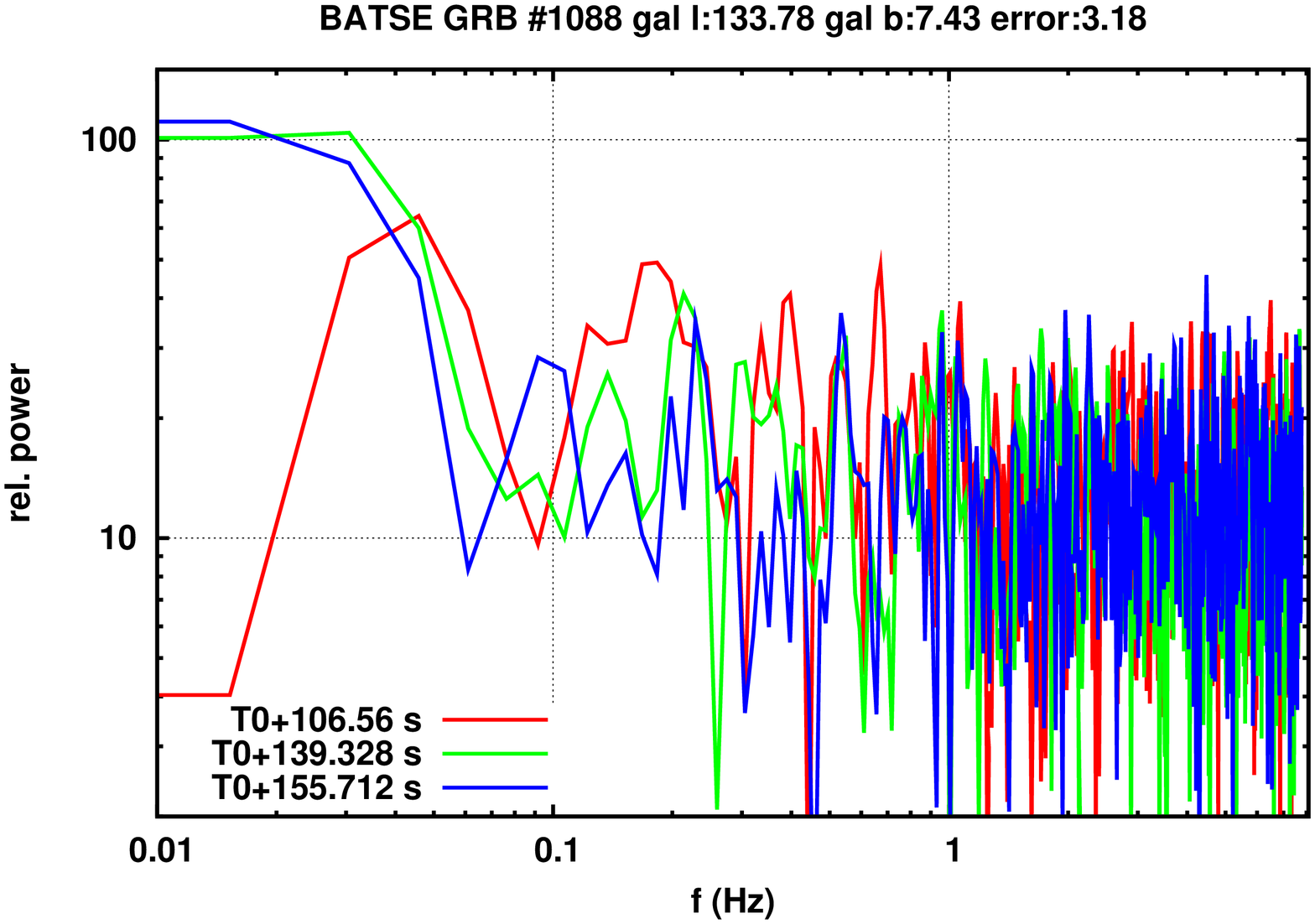}
\includegraphics[height=.656\columnwidth, angle=270]{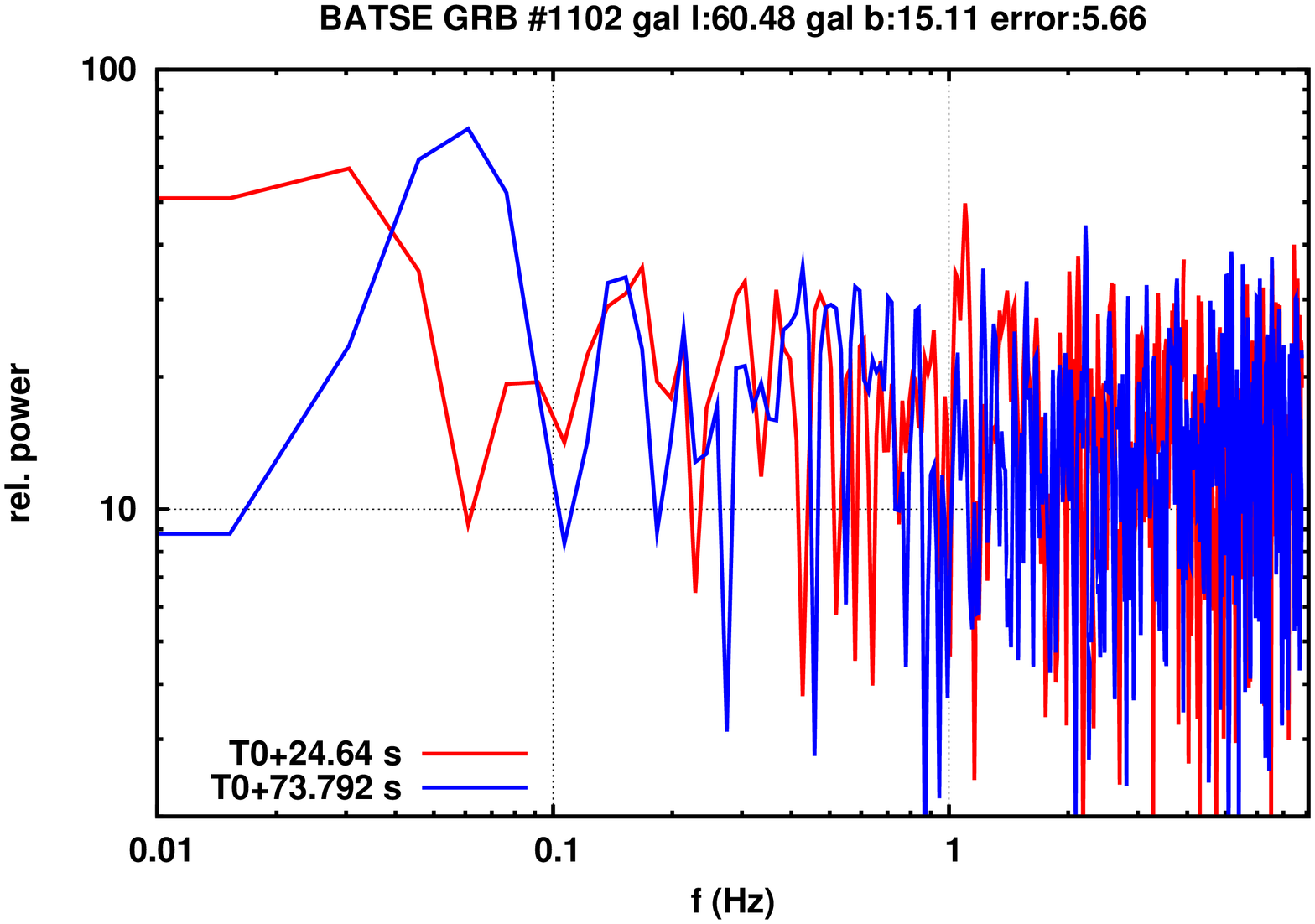}

\includegraphics[height=.656\columnwidth, angle=270]{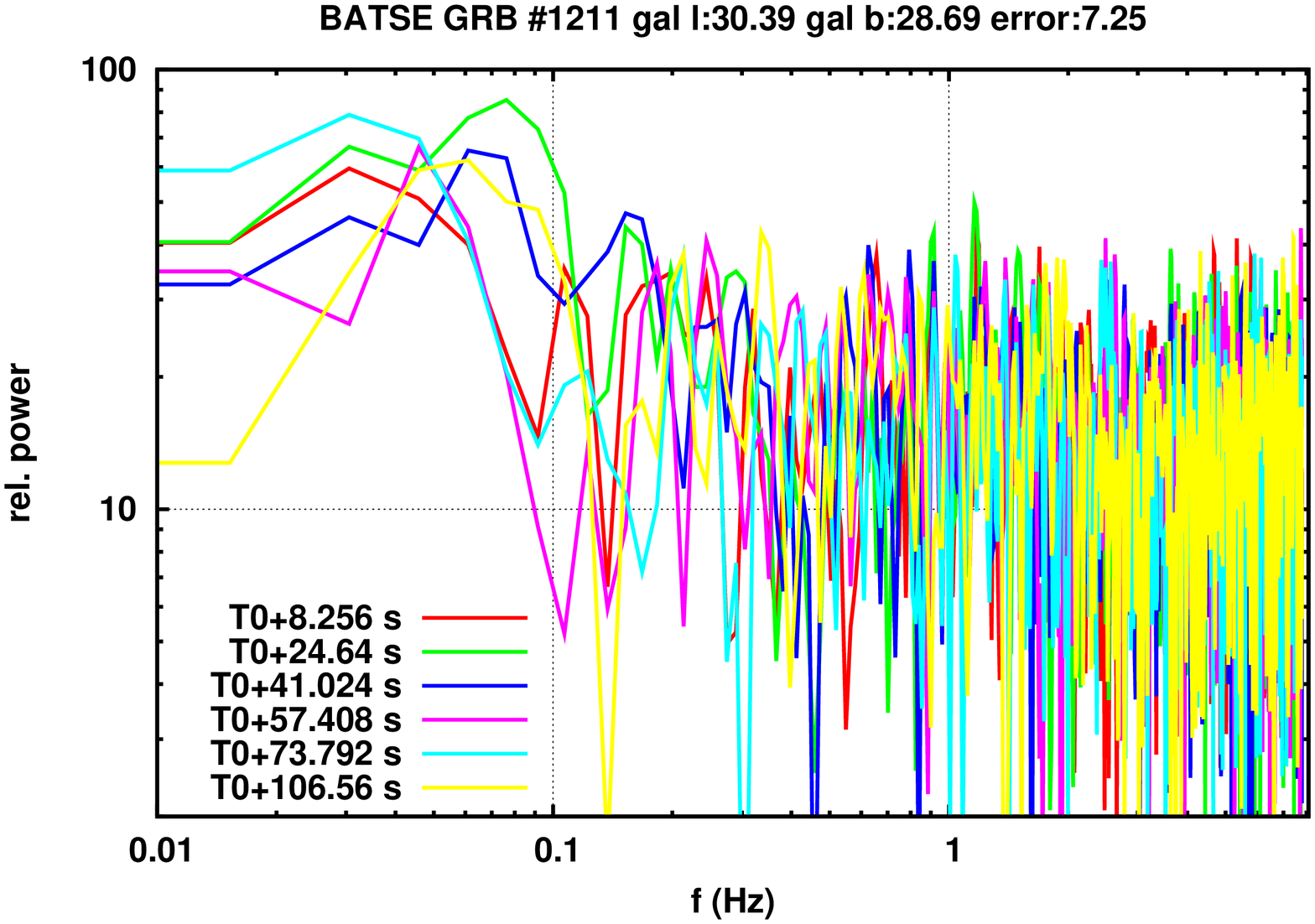}
\includegraphics[height=.656\columnwidth, angle=270]{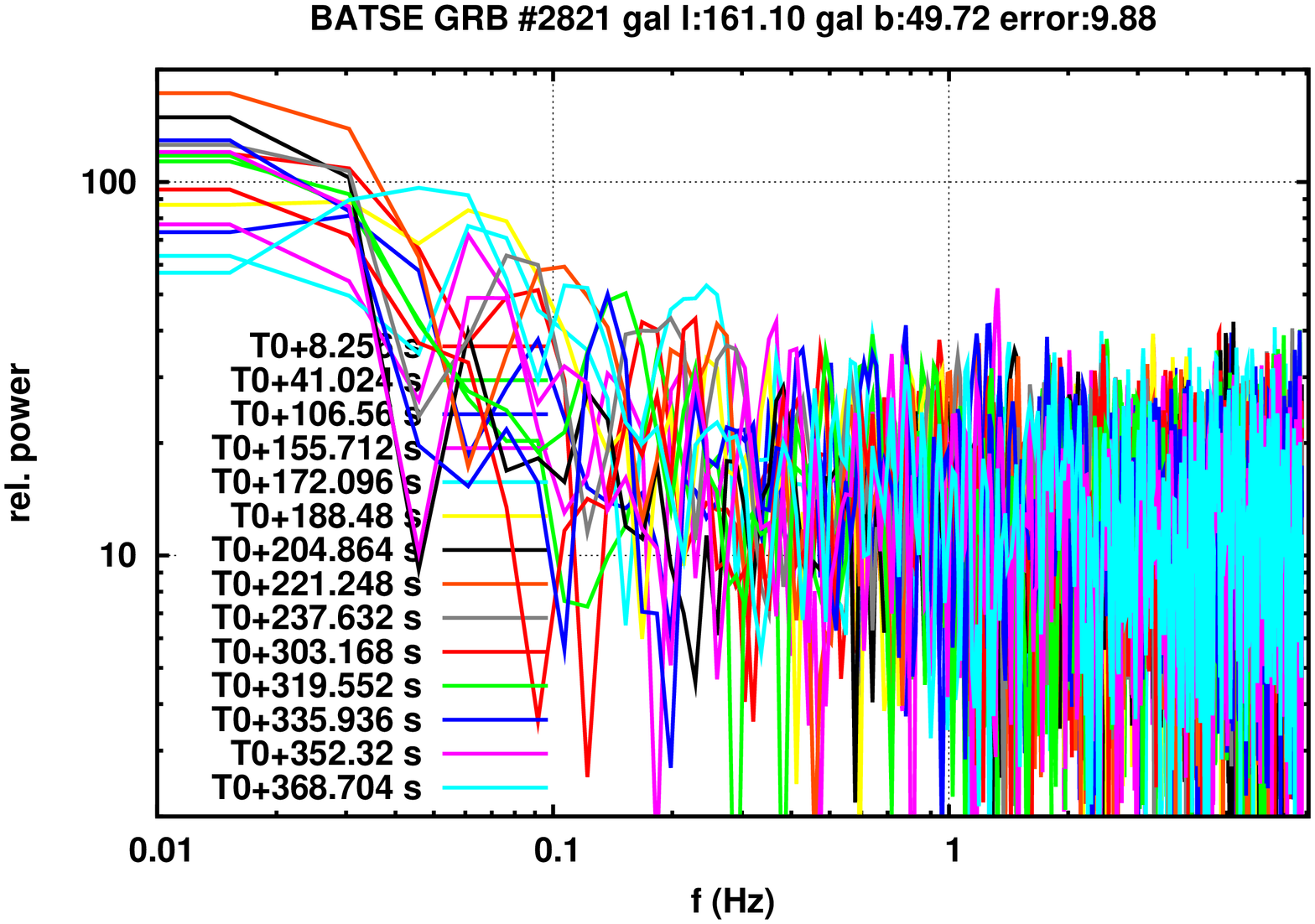}

\includegraphics[height=.656\columnwidth, angle=270]{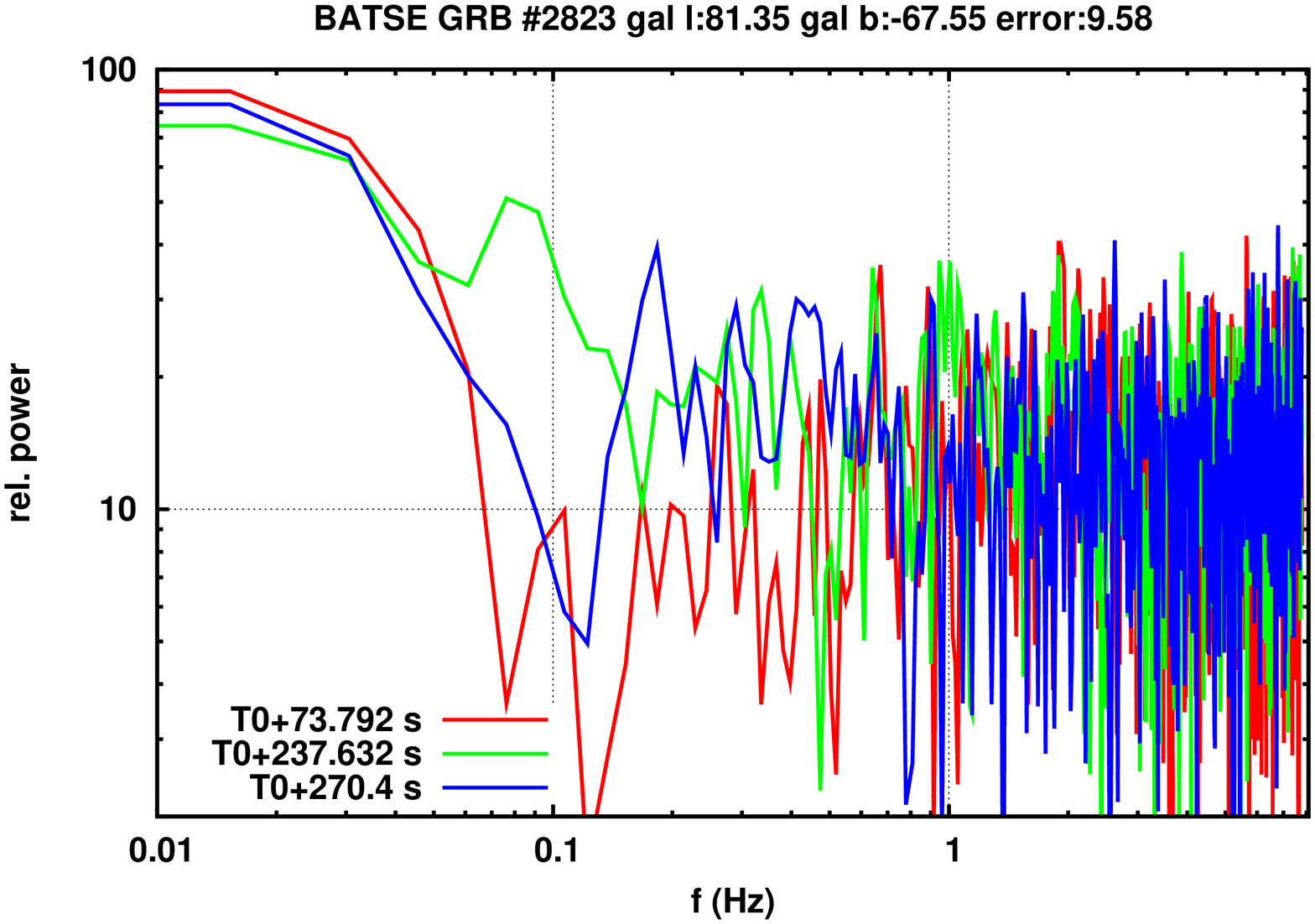}
\includegraphics[height=.656\columnwidth, angle=270]{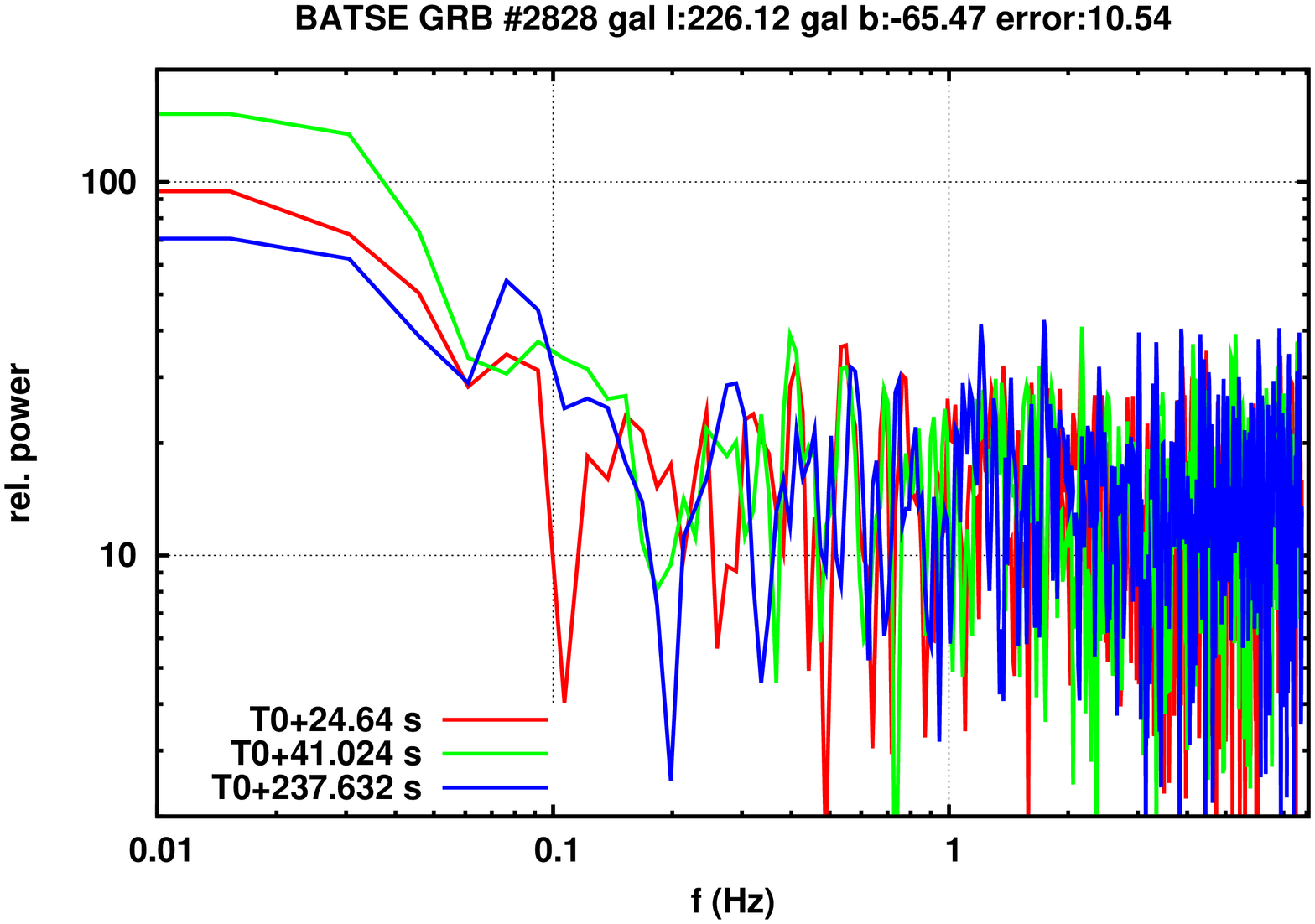}

\includegraphics[height=.656\columnwidth, angle=270]{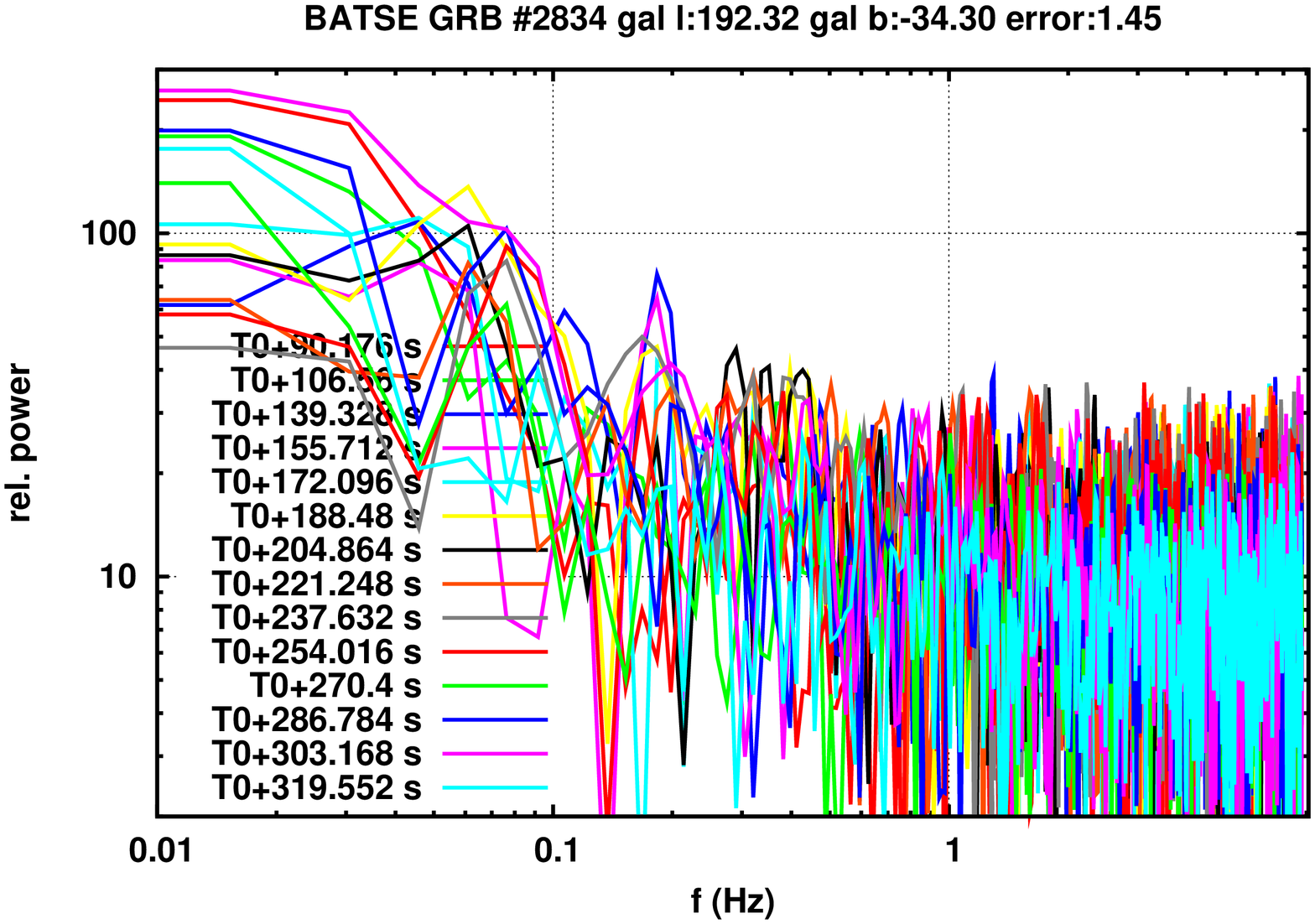}
\includegraphics[height=.656\columnwidth, angle=270]{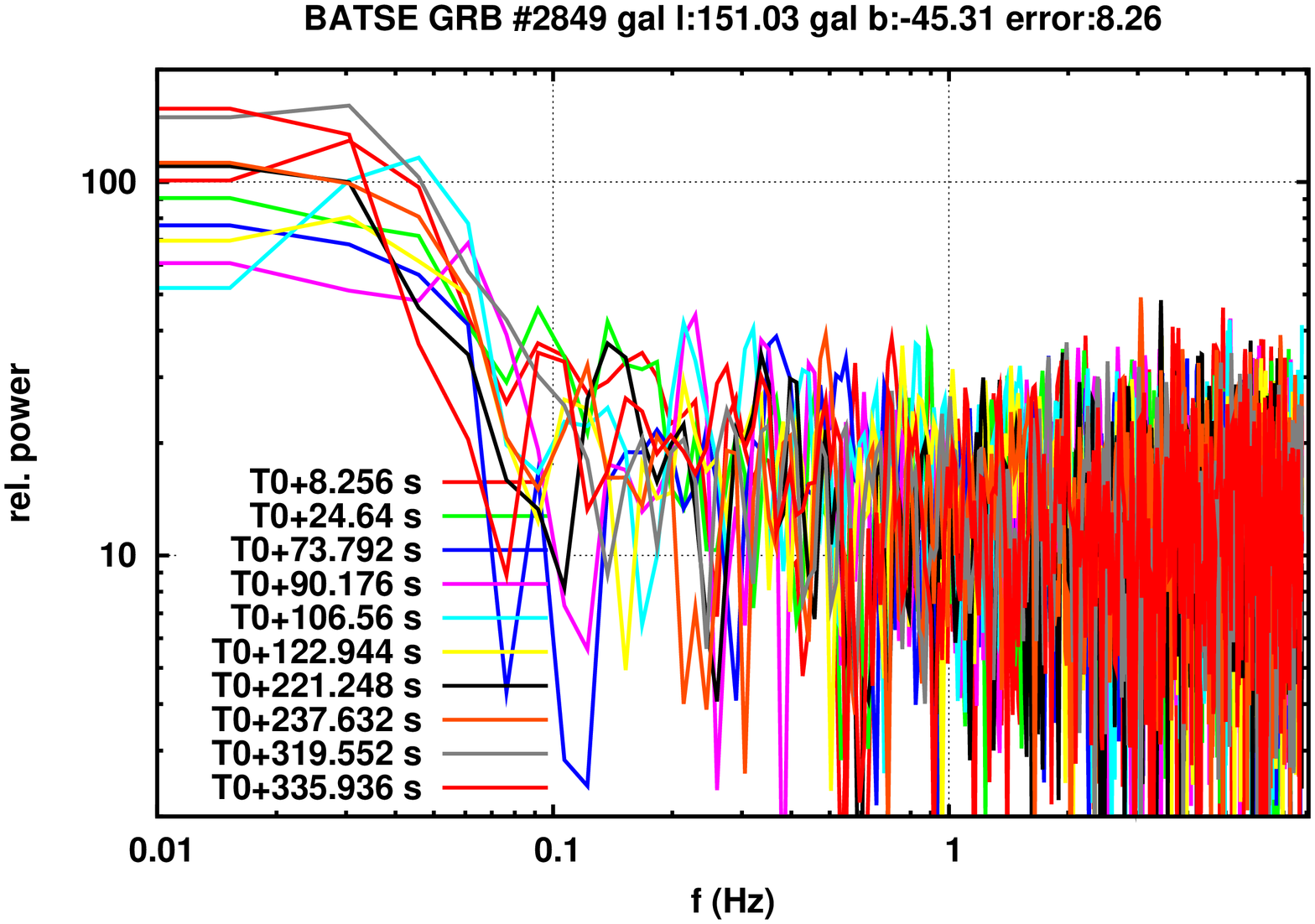}
\caption{Unusually high power spectra of the selected BATSE triggers with different window start points.}
\end{figure*}

\begin{figure*}[t]
          \centering
\includegraphics[height=.656\columnwidth, angle=270]{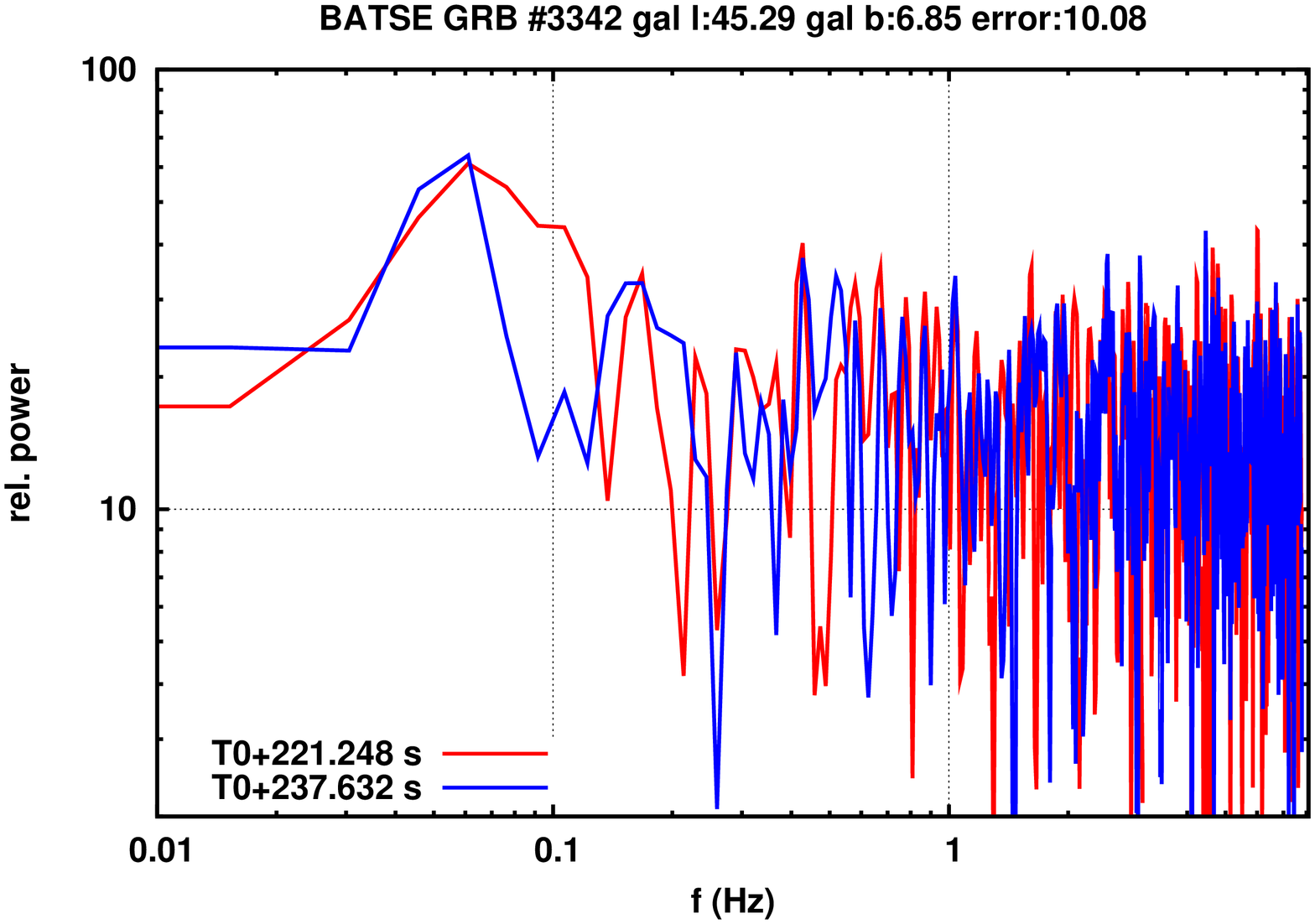}
\includegraphics[height=.656\columnwidth, angle=270]{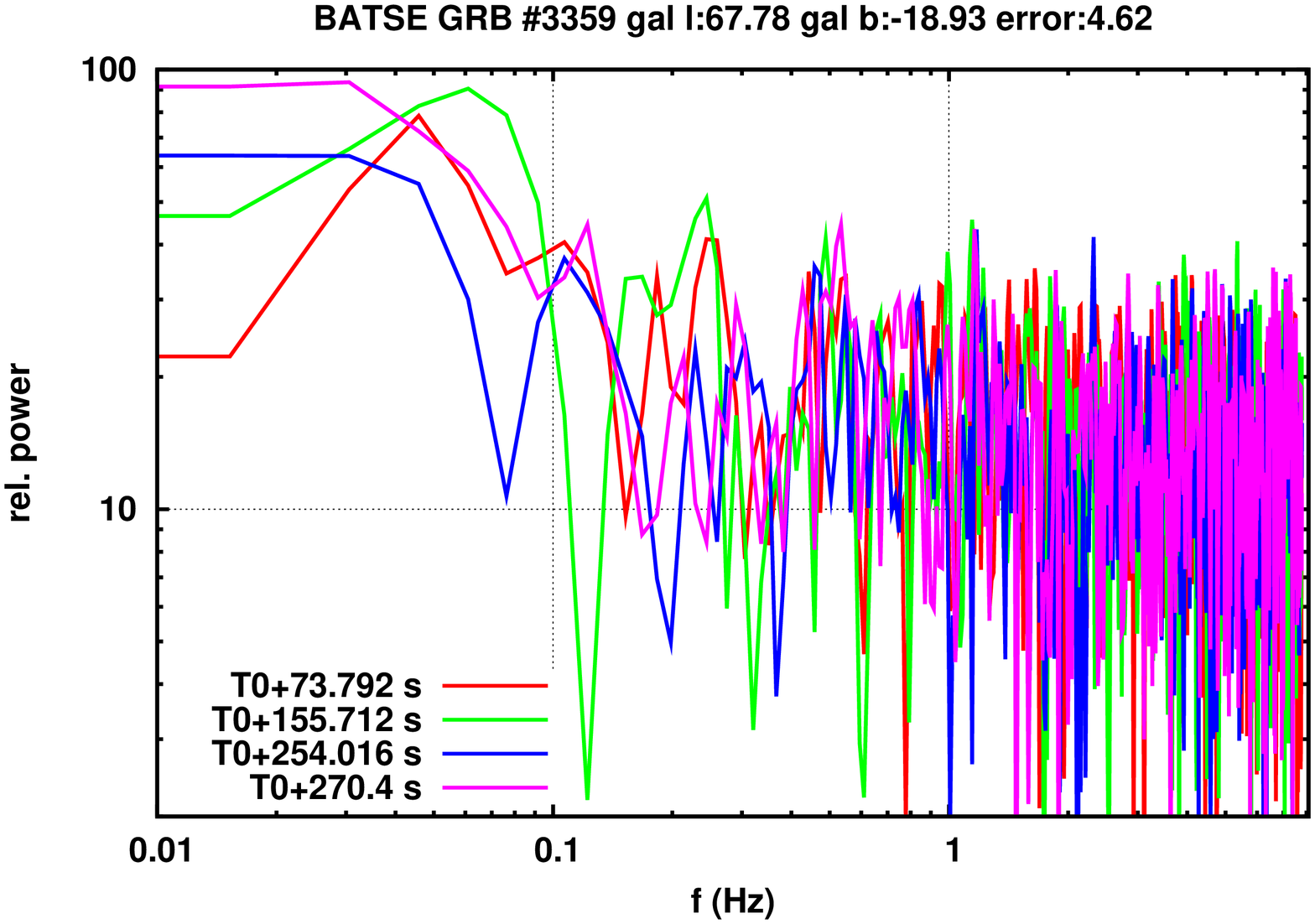}

\includegraphics[height=.656\columnwidth, angle=270]{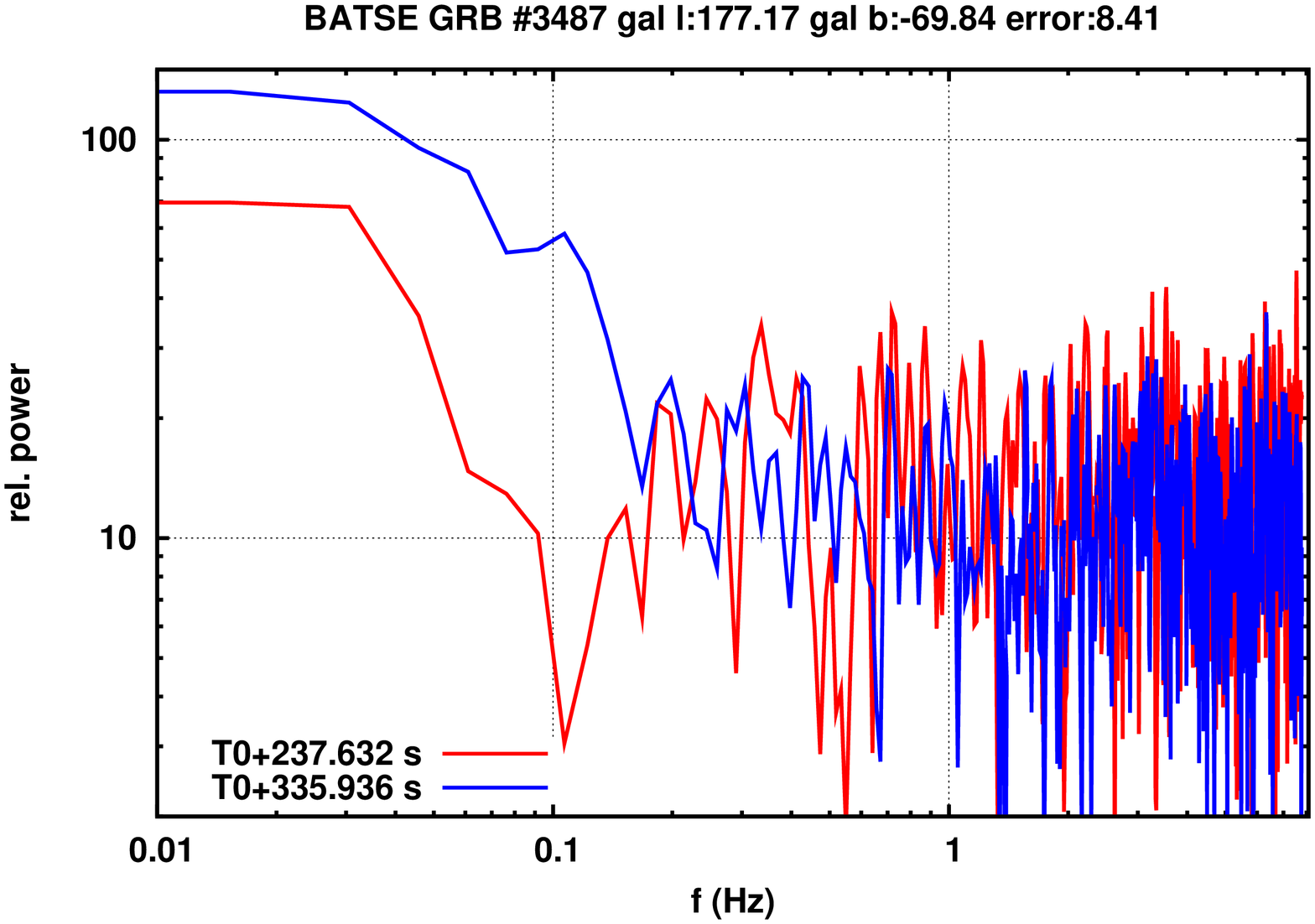}
\includegraphics[height=.656\columnwidth, angle=270]{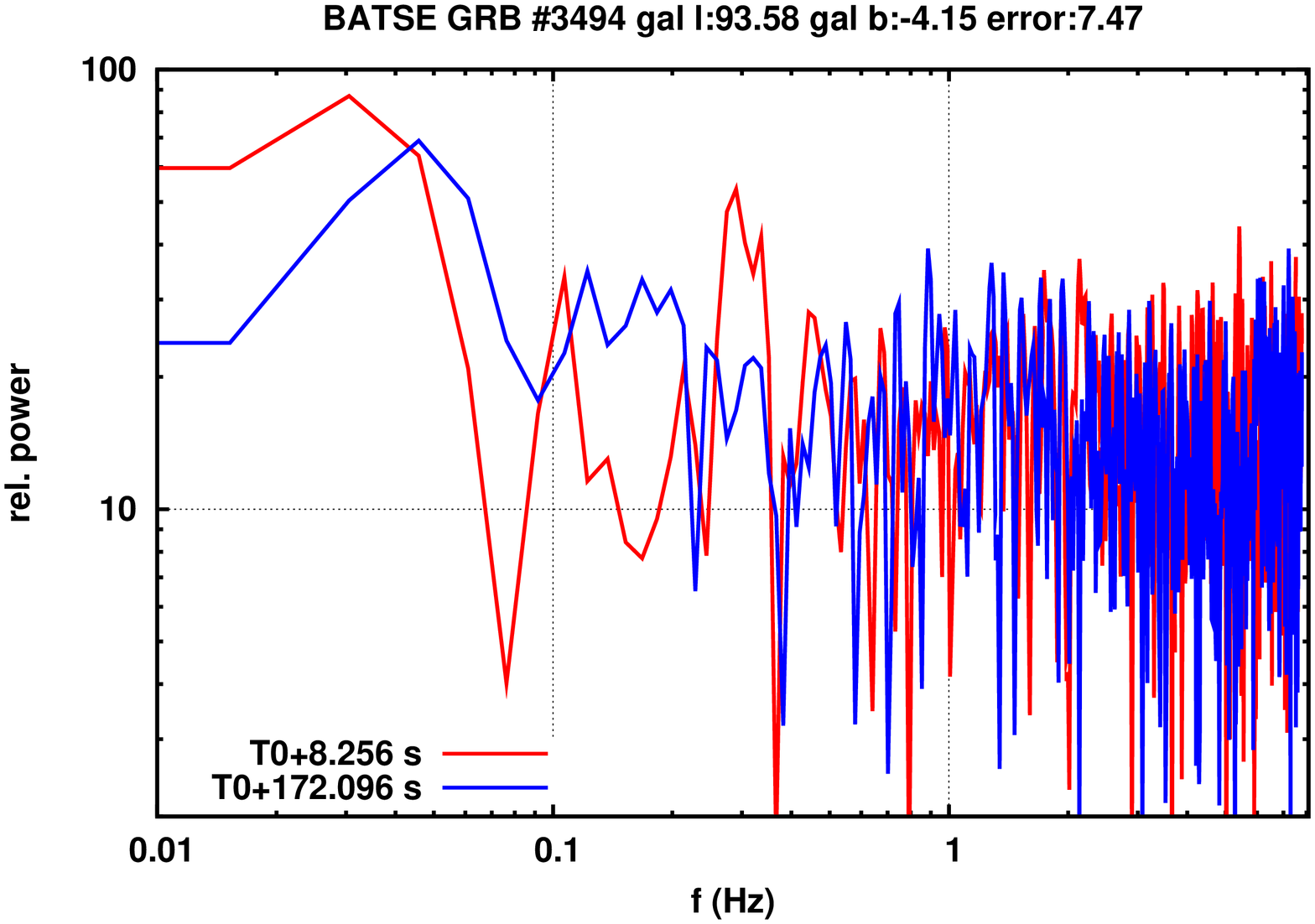}

\includegraphics[height=.656\columnwidth, angle=270]{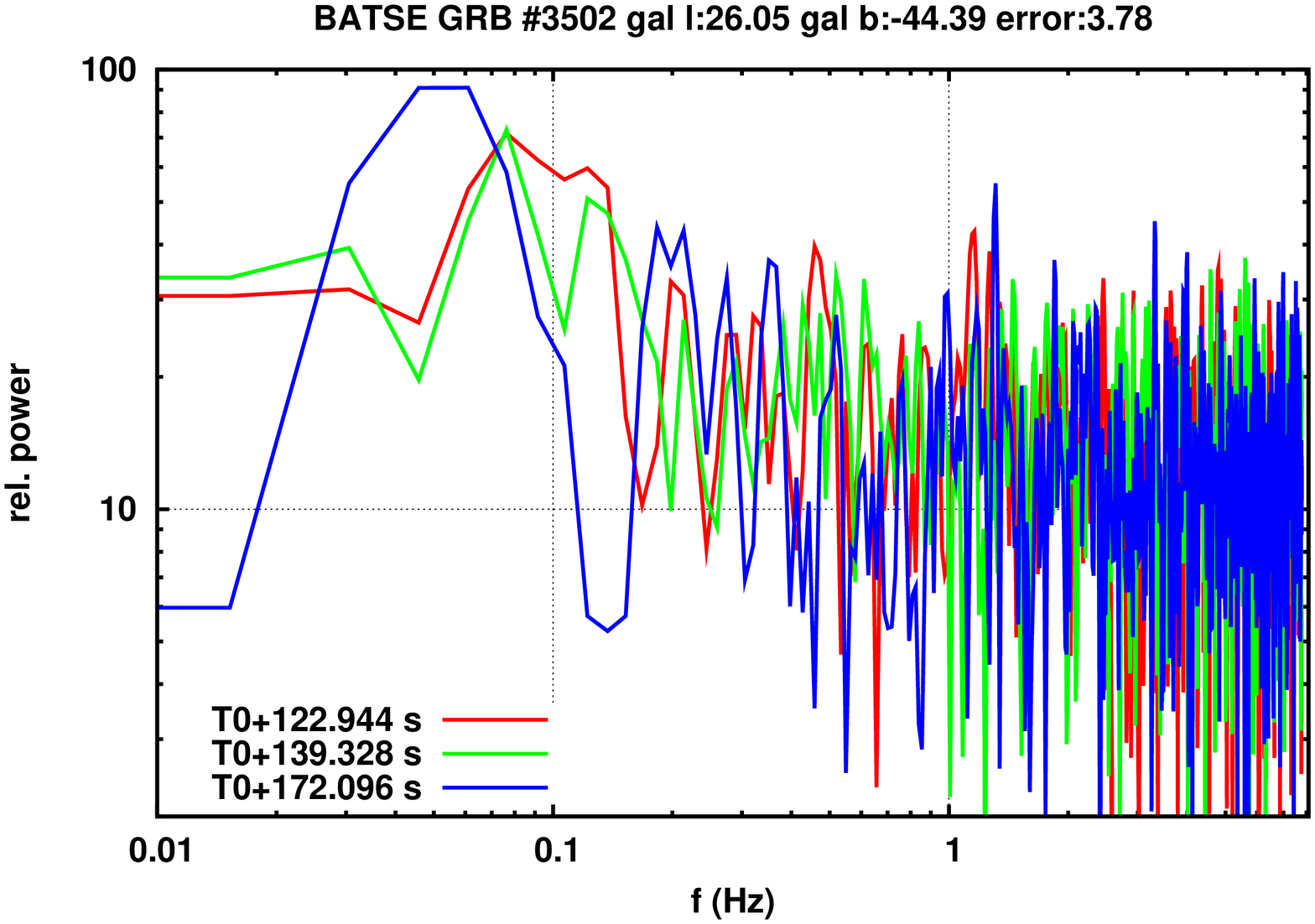}
\includegraphics[height=.656\columnwidth, angle=270]{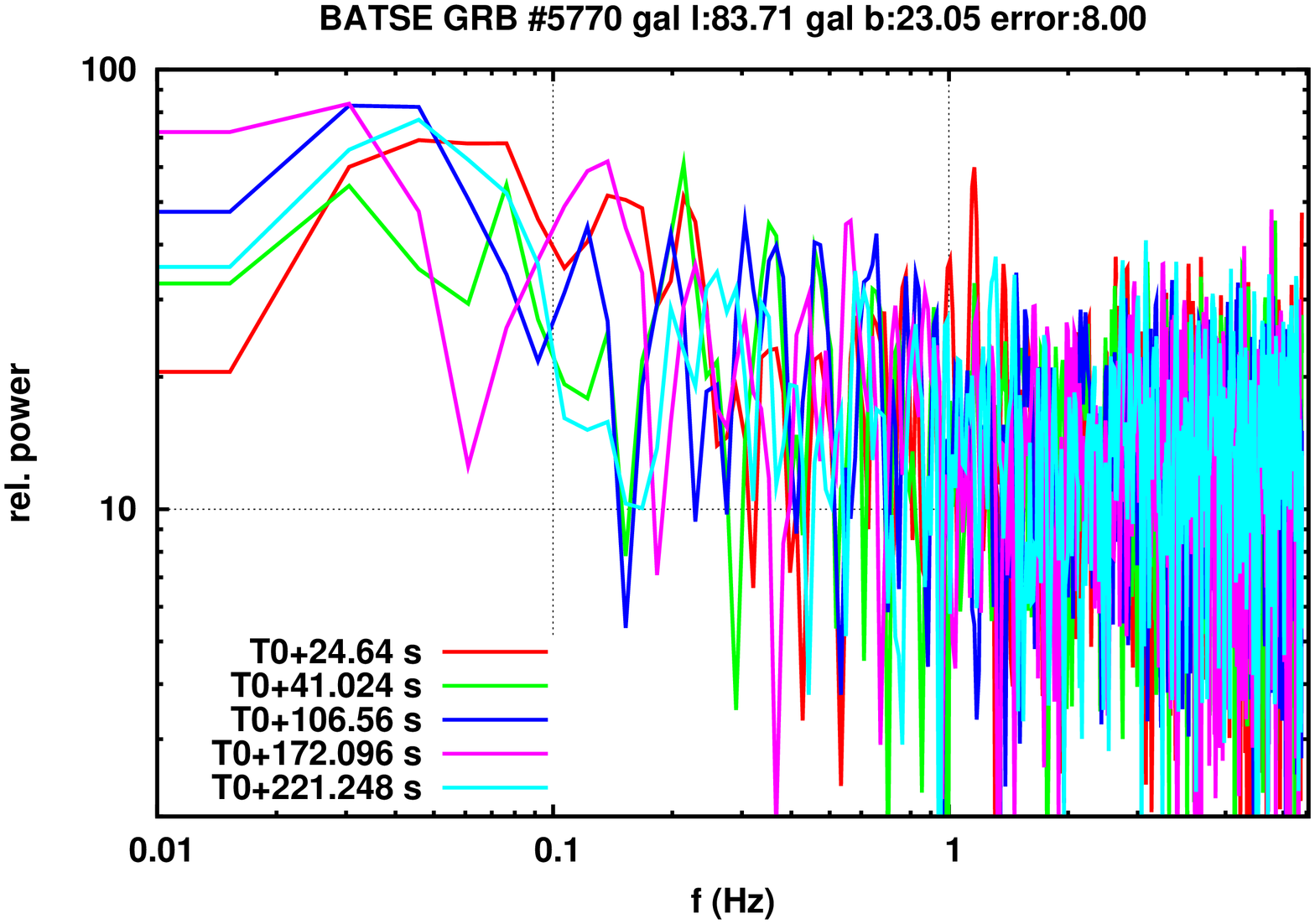}

\includegraphics[height=.656\columnwidth, angle=270]{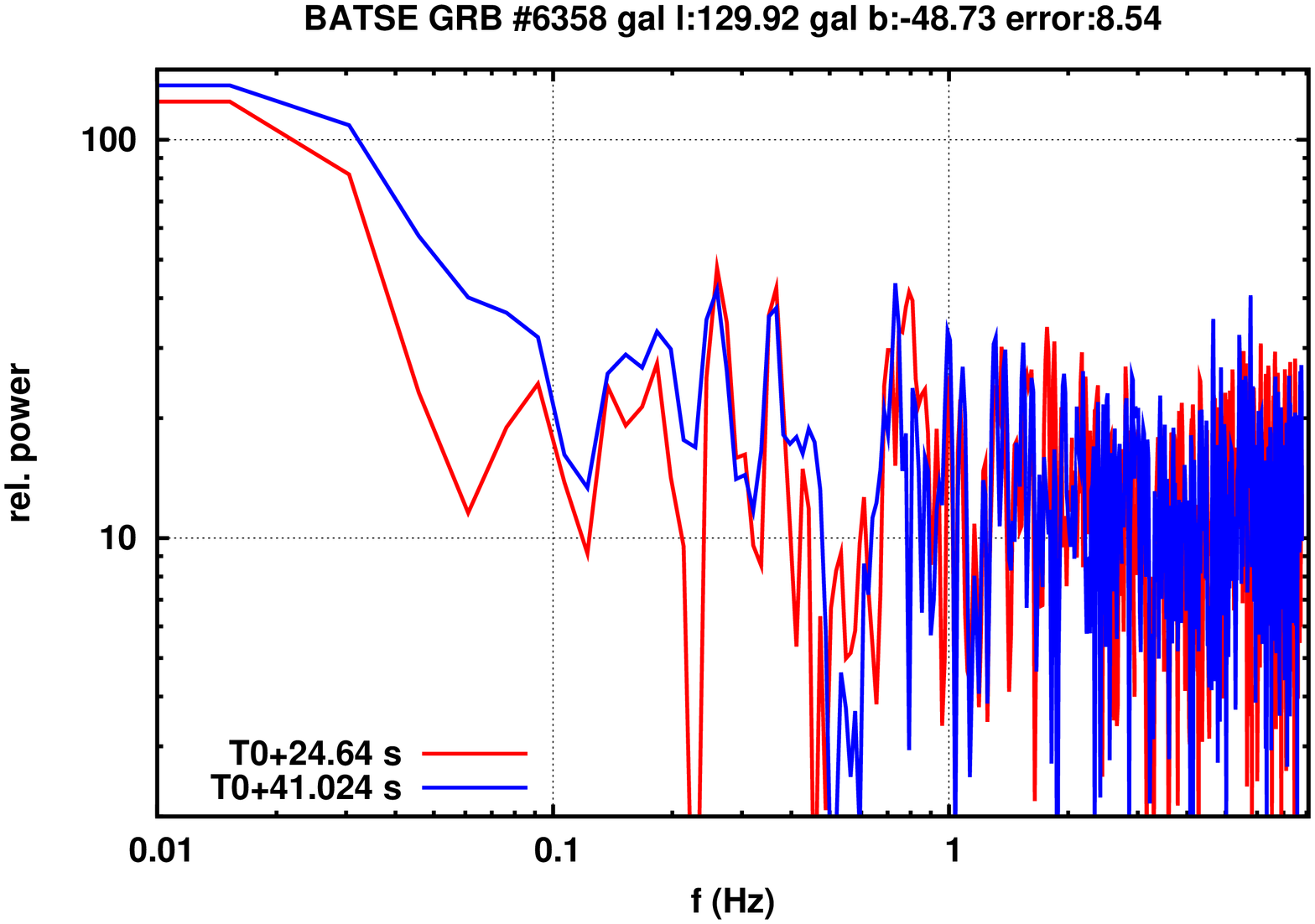}
\includegraphics[height=.656\columnwidth, angle=270]{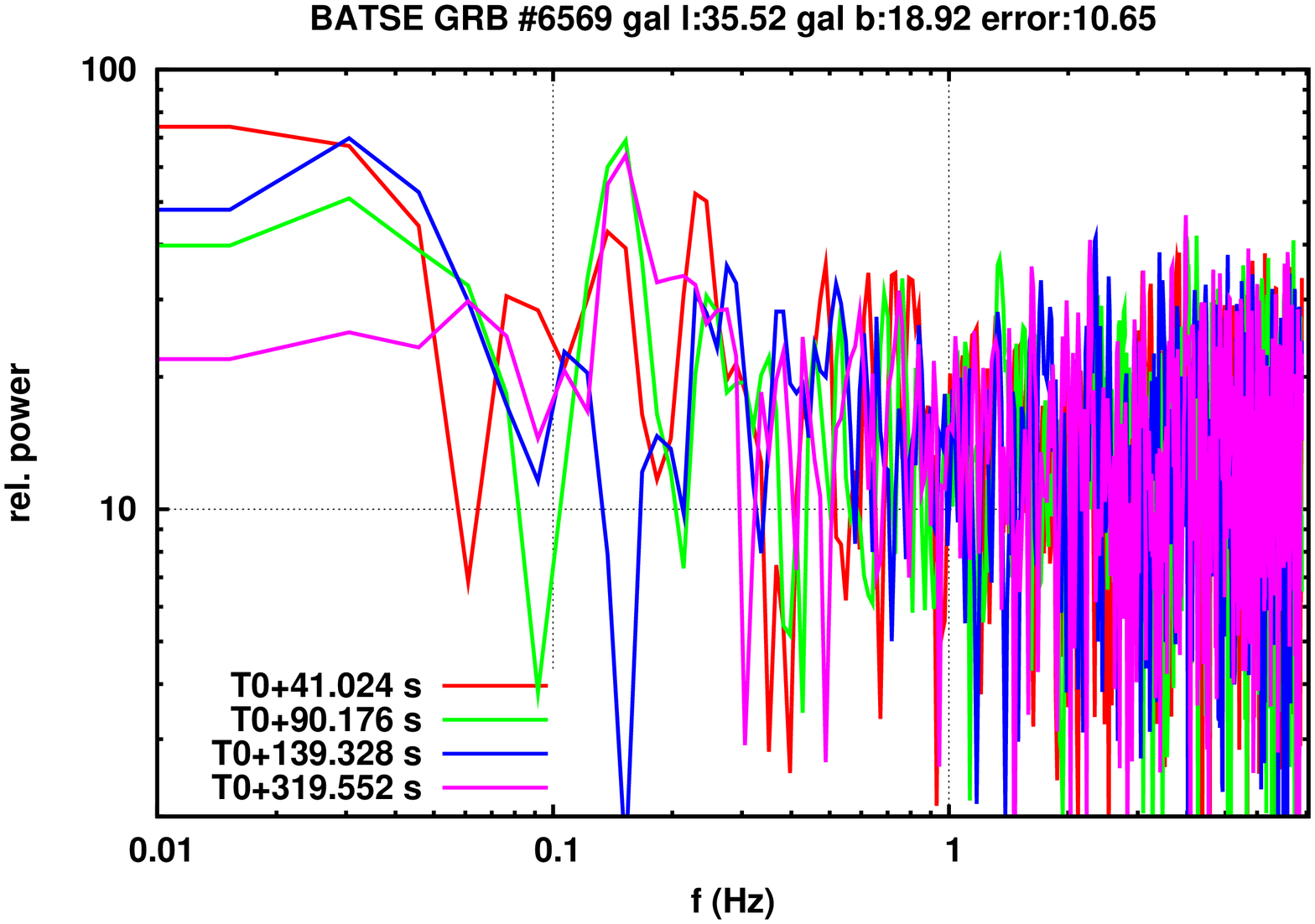}

\includegraphics[height=.656\columnwidth, angle=270]{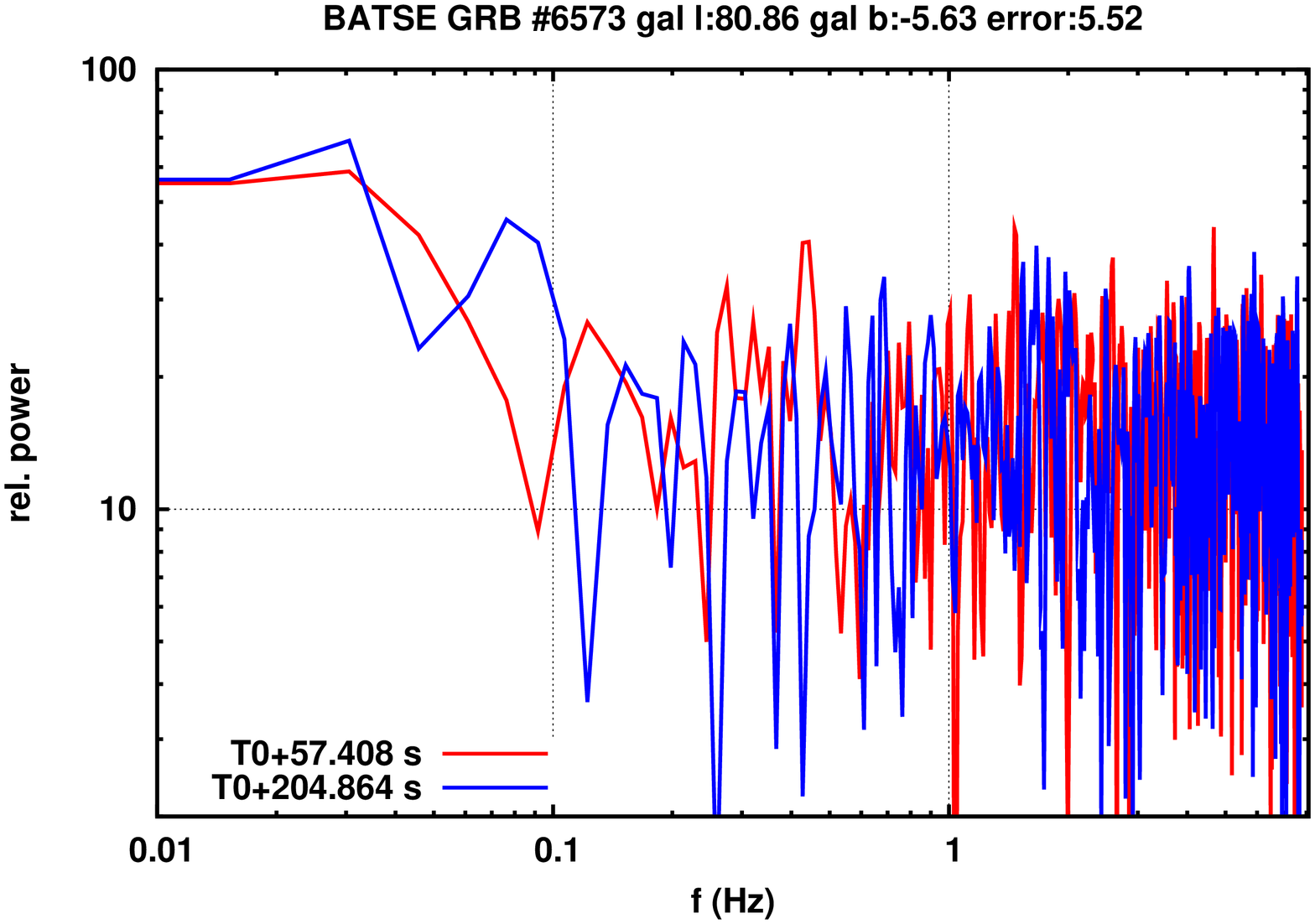}
\includegraphics[height=.656\columnwidth, angle=270]{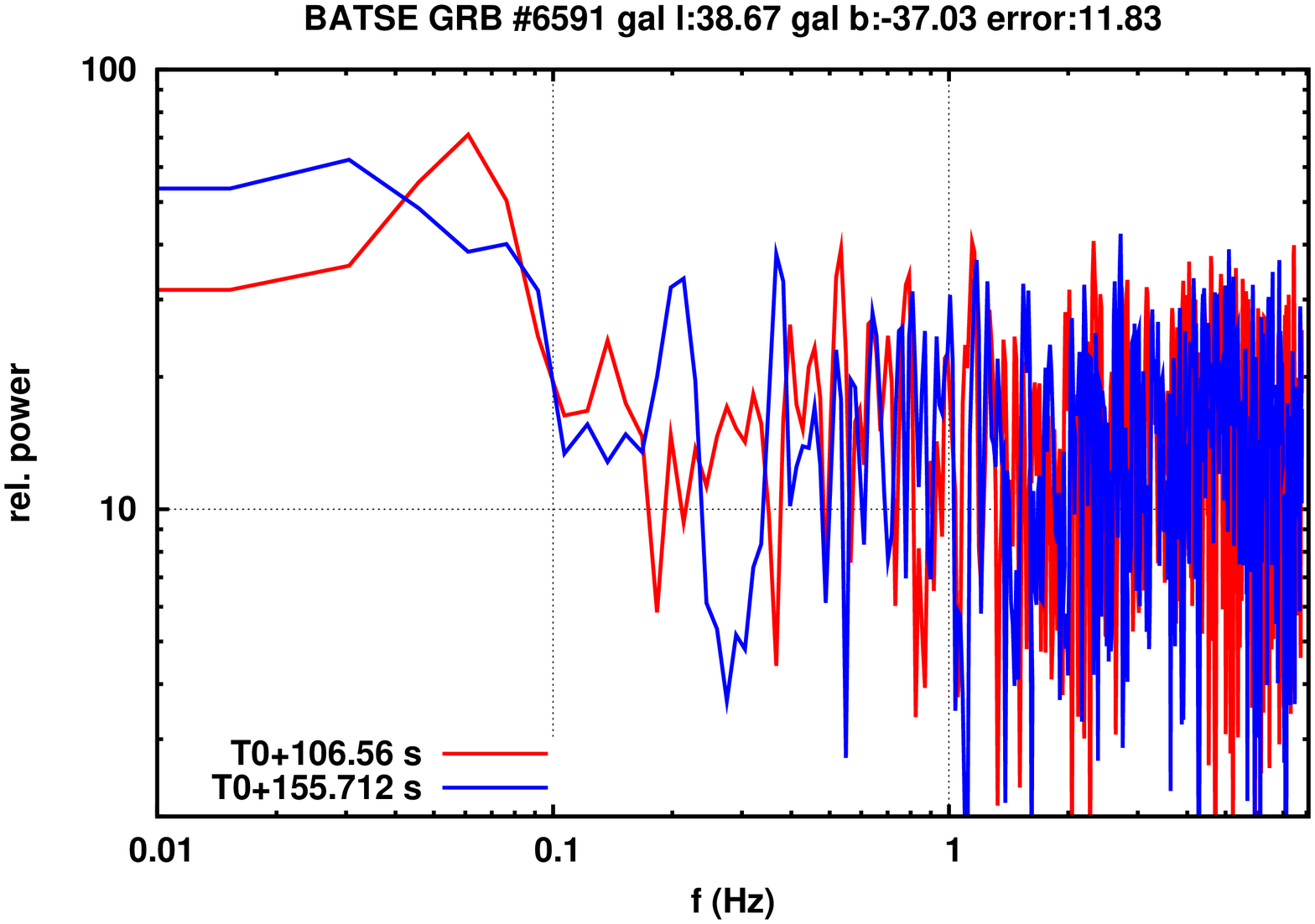}
\bigskip\vspace{0.2cm}
\caption{FIG. 1 (continued): Unusually high power spectra of the selected BATSE triggers with different window start points.}
\end{figure*}

\section{Pulse shapes} 

On Fig. 2. we show some triggers' folded phase-frequency diagram and (selected)
folded pulse shape.  One can observe that some of the signals appears to be
periodic: however we should keep in mind that for these triggers have only one
kind of observation with fixed time resolution ($64 ms$), so it is very hard to
assign real significance levels to these curves. 

Triggers \#0373 and \#6591 show a broad signal, while \#0434 and \#3359 show
more narrow peaks.  It is remarkable that all suspicious frequencies are beyond
any known BATSE periodic detector effect.

           \begin{figure*}[t]\centering
\includegraphics[height=.75\columnwidth, angle=270]{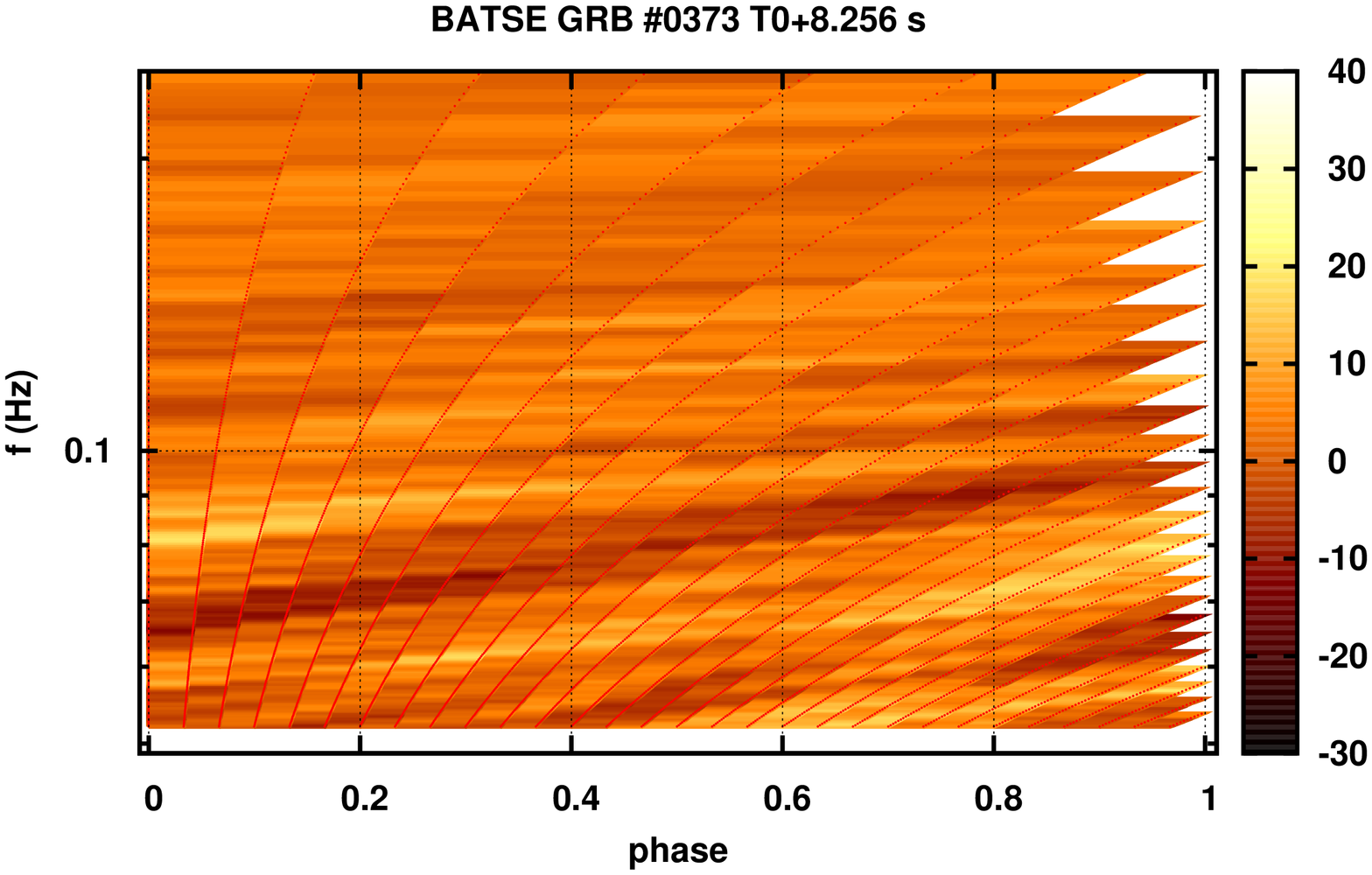}\includegraphics[height=.75\columnwidth, angle=270]{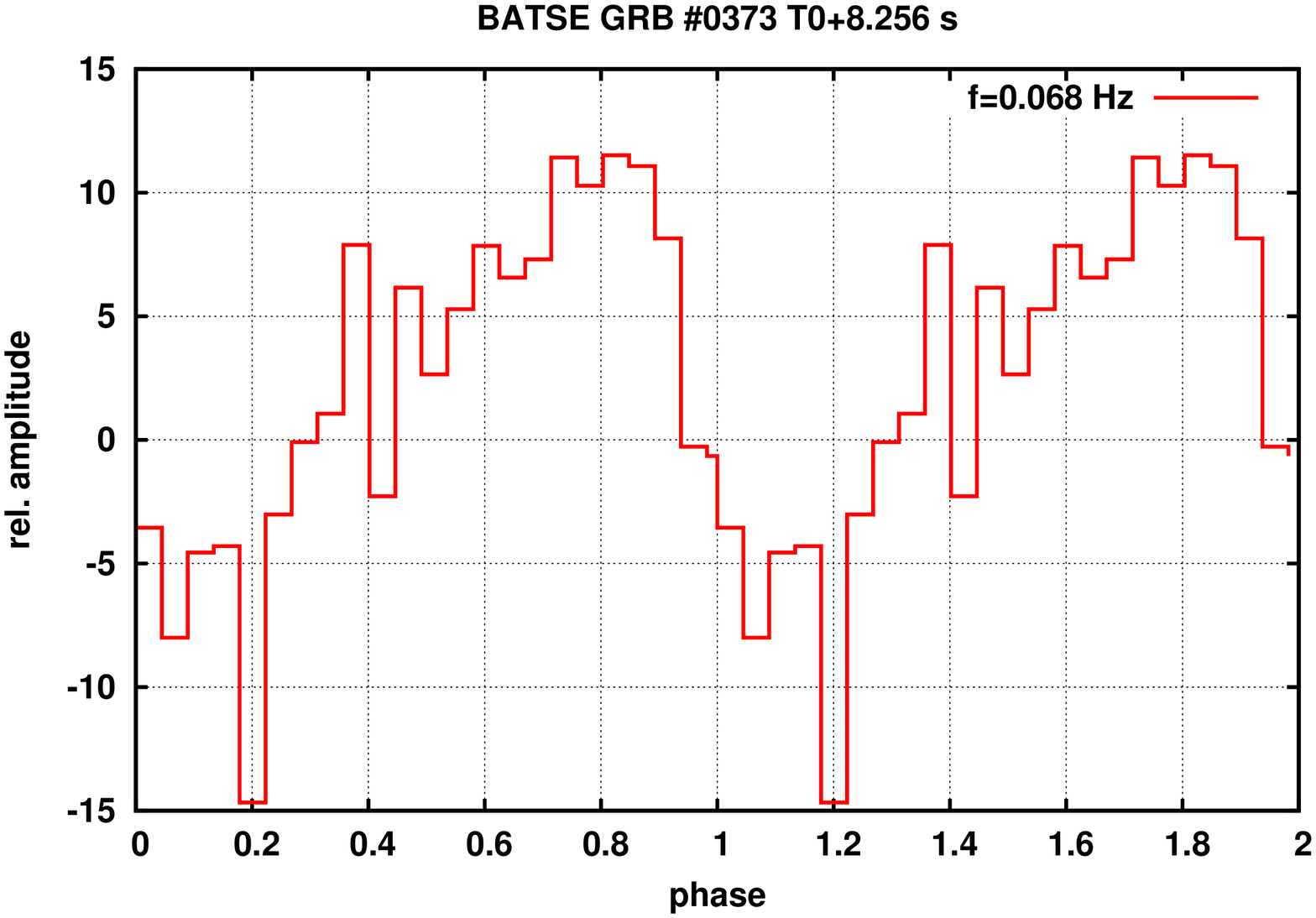}

\includegraphics[height=.75\columnwidth, angle=270]{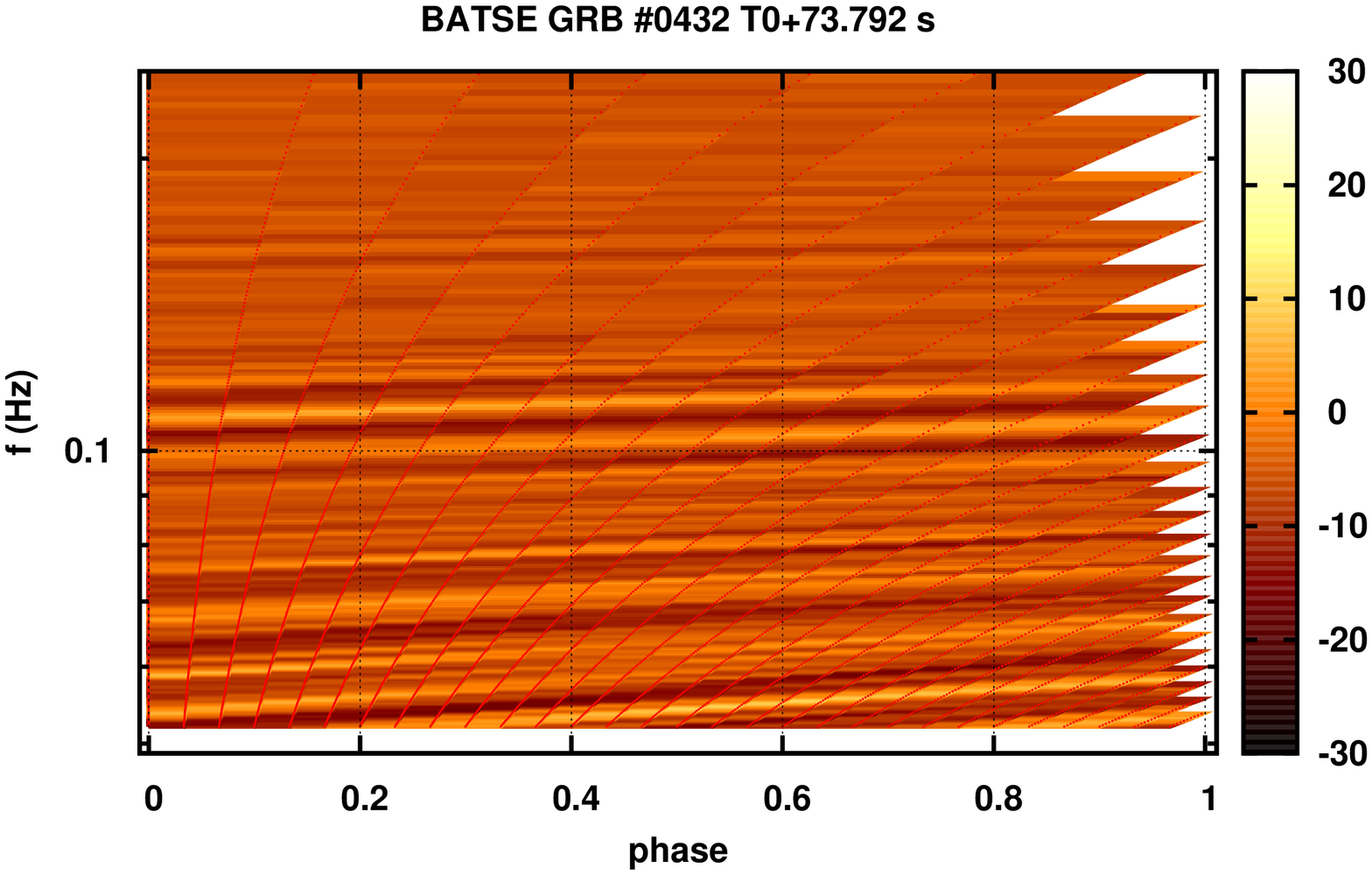} \includegraphics[height=.75\columnwidth, angle=270]{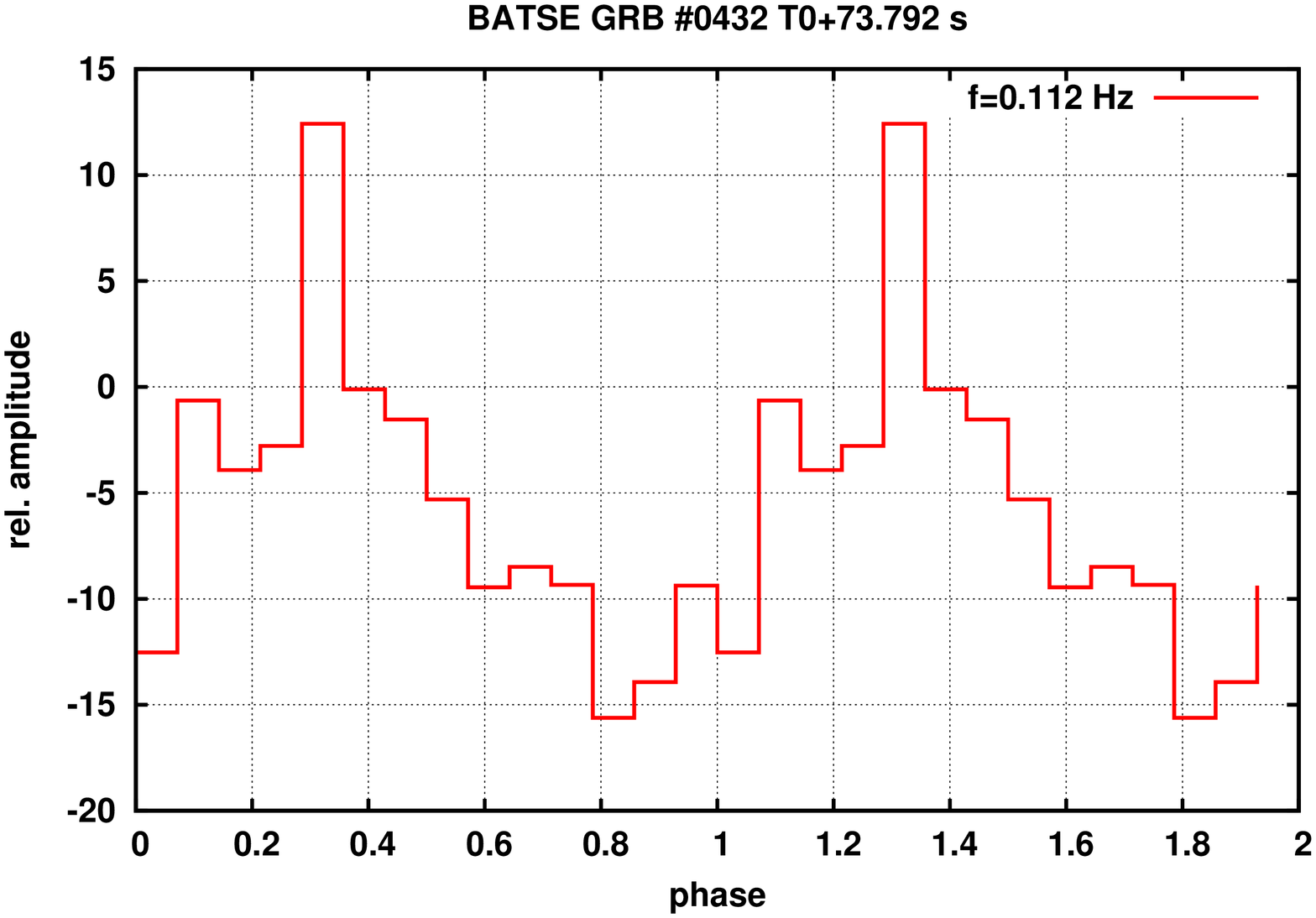}

\includegraphics[height=.75\columnwidth, angle=270]{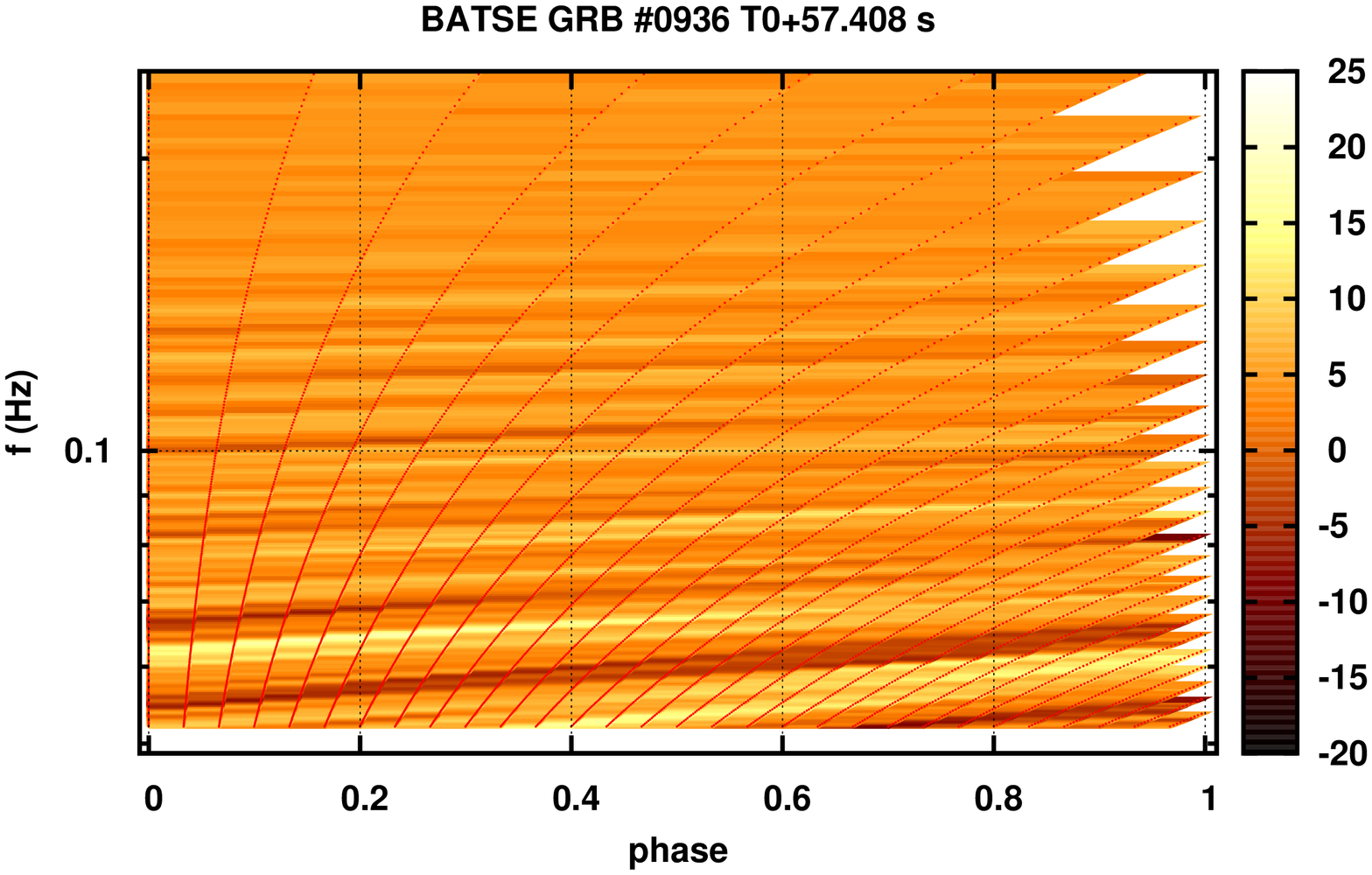} \includegraphics[height=.75\columnwidth, angle=270]{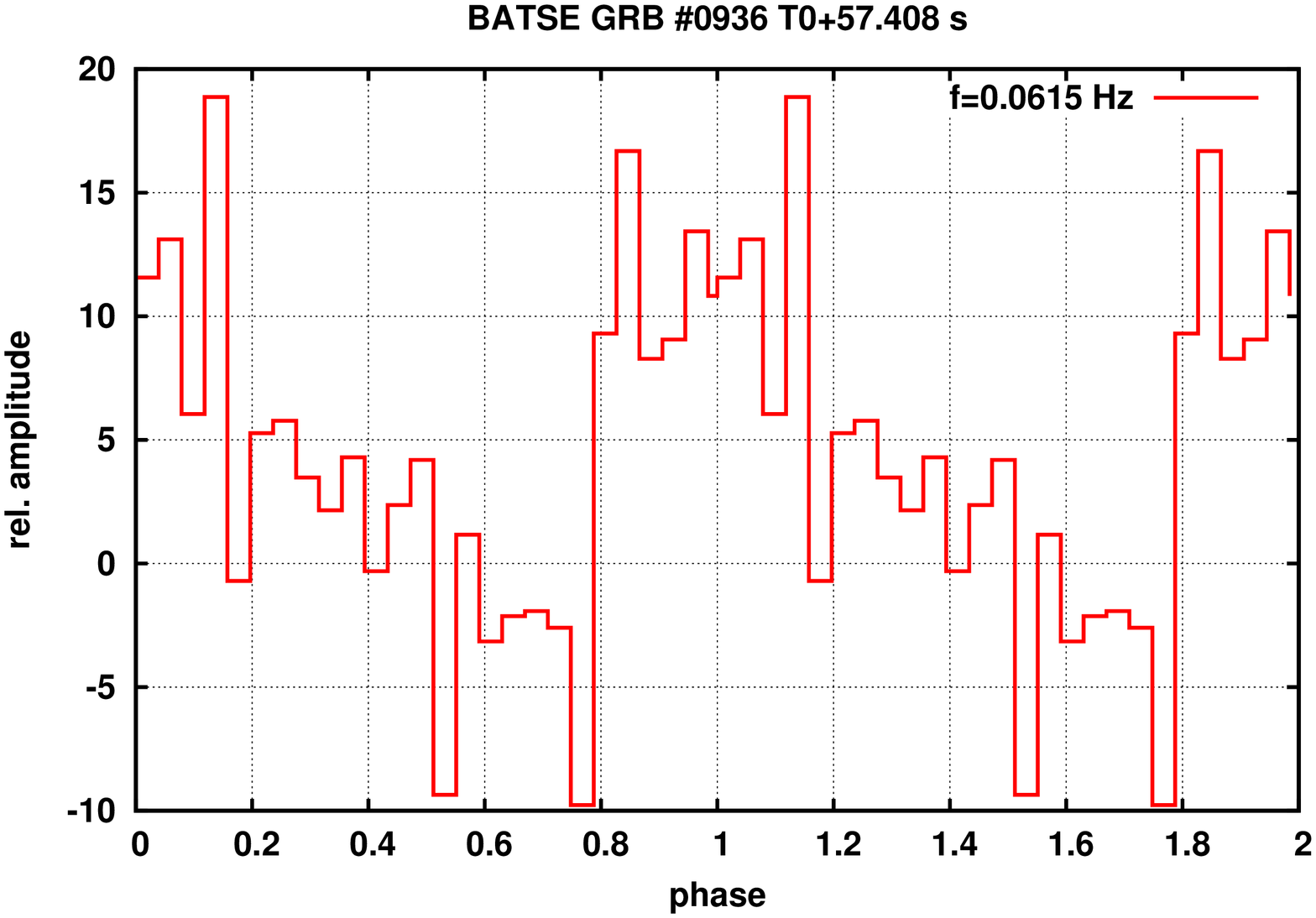}

\includegraphics[height=.75\columnwidth, angle=270]{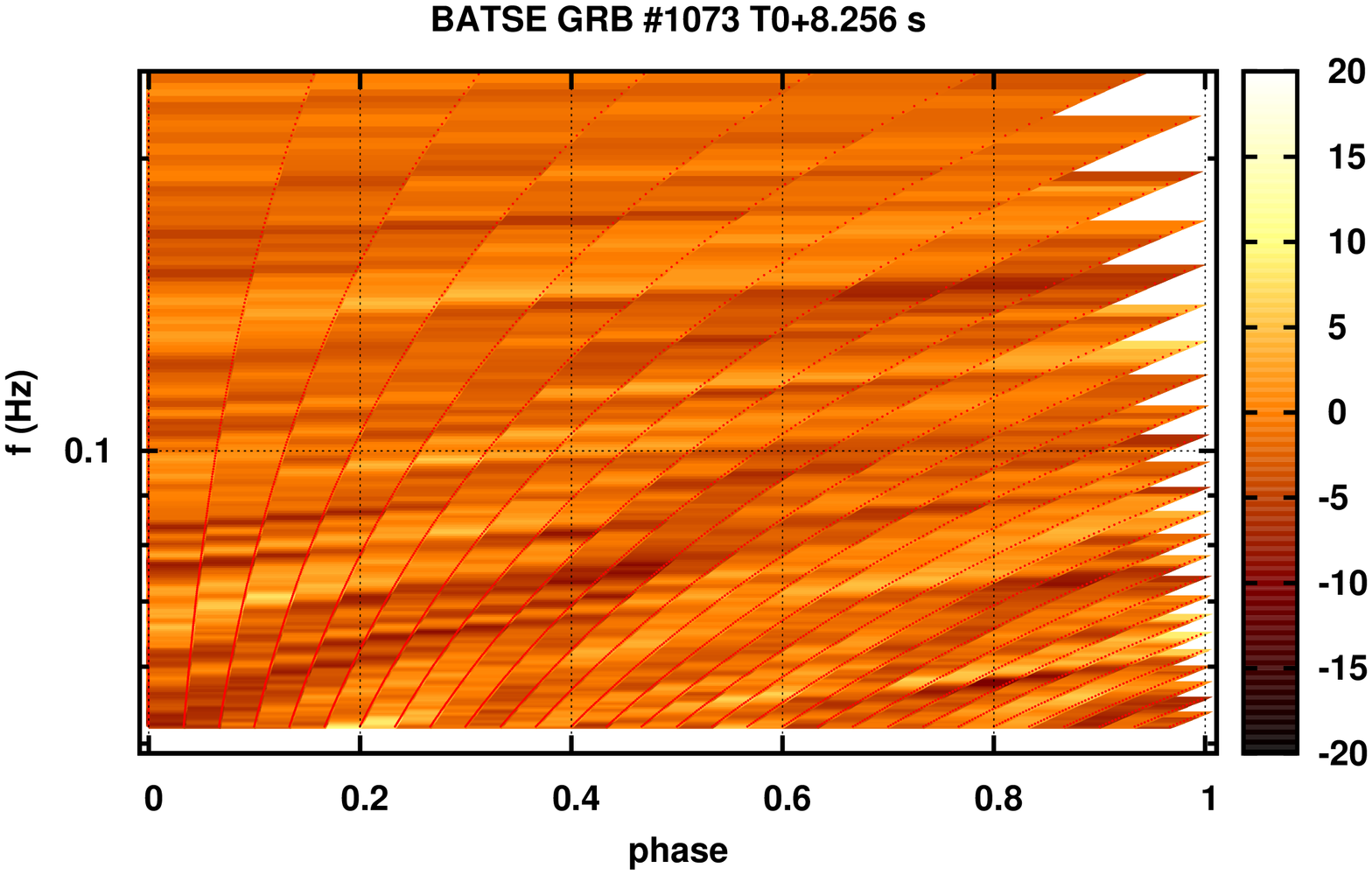} \includegraphics[height=.75\columnwidth, angle=270]{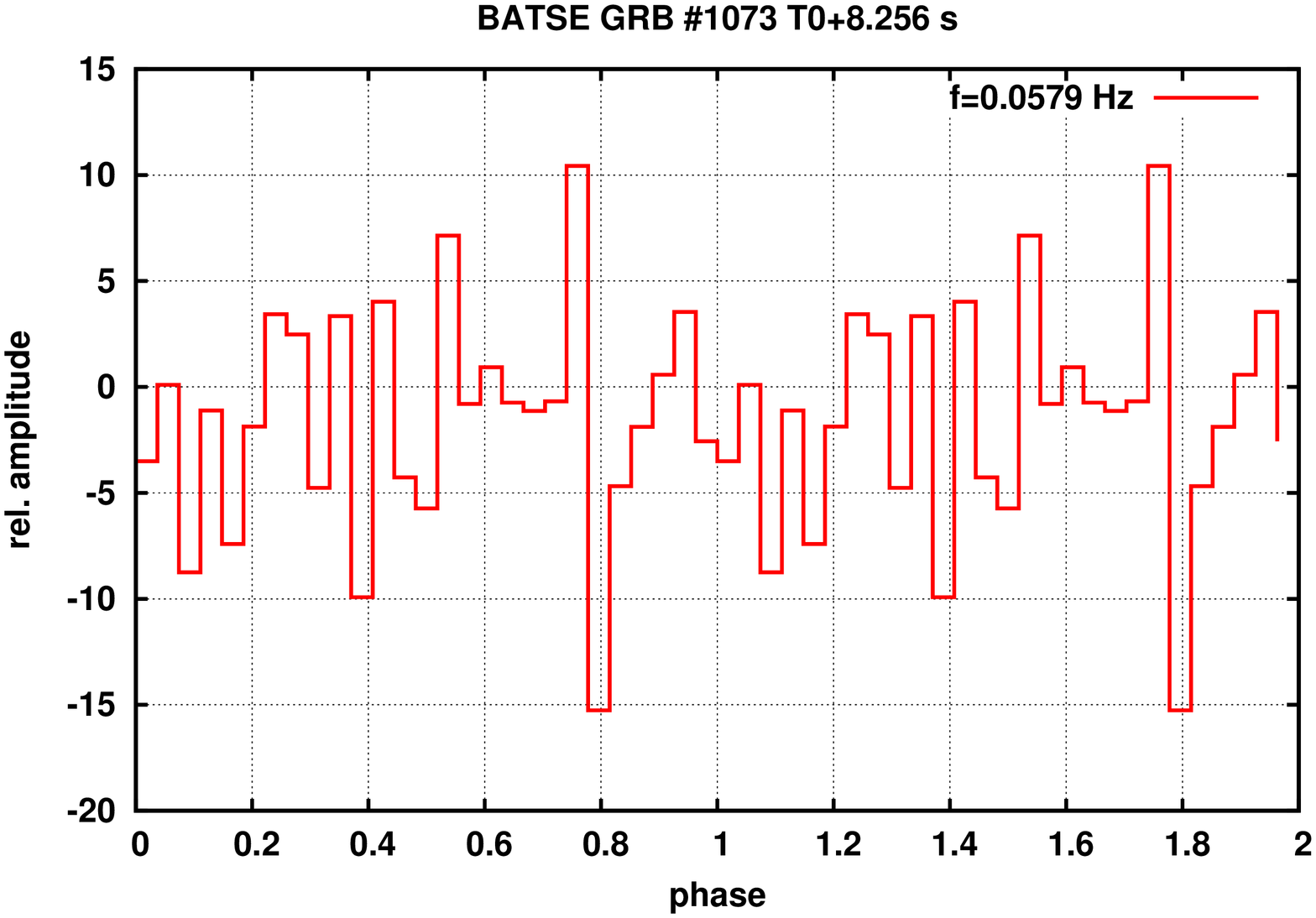}

\bigskip\vspace{0.2cm}
\caption{Some folded phase-frequency diagrams and folded pulse shapes of the selected triggers.  }
           \end{figure*}

           \begin{figure*}[t]\centering
\includegraphics[height=.75\columnwidth, angle=270]{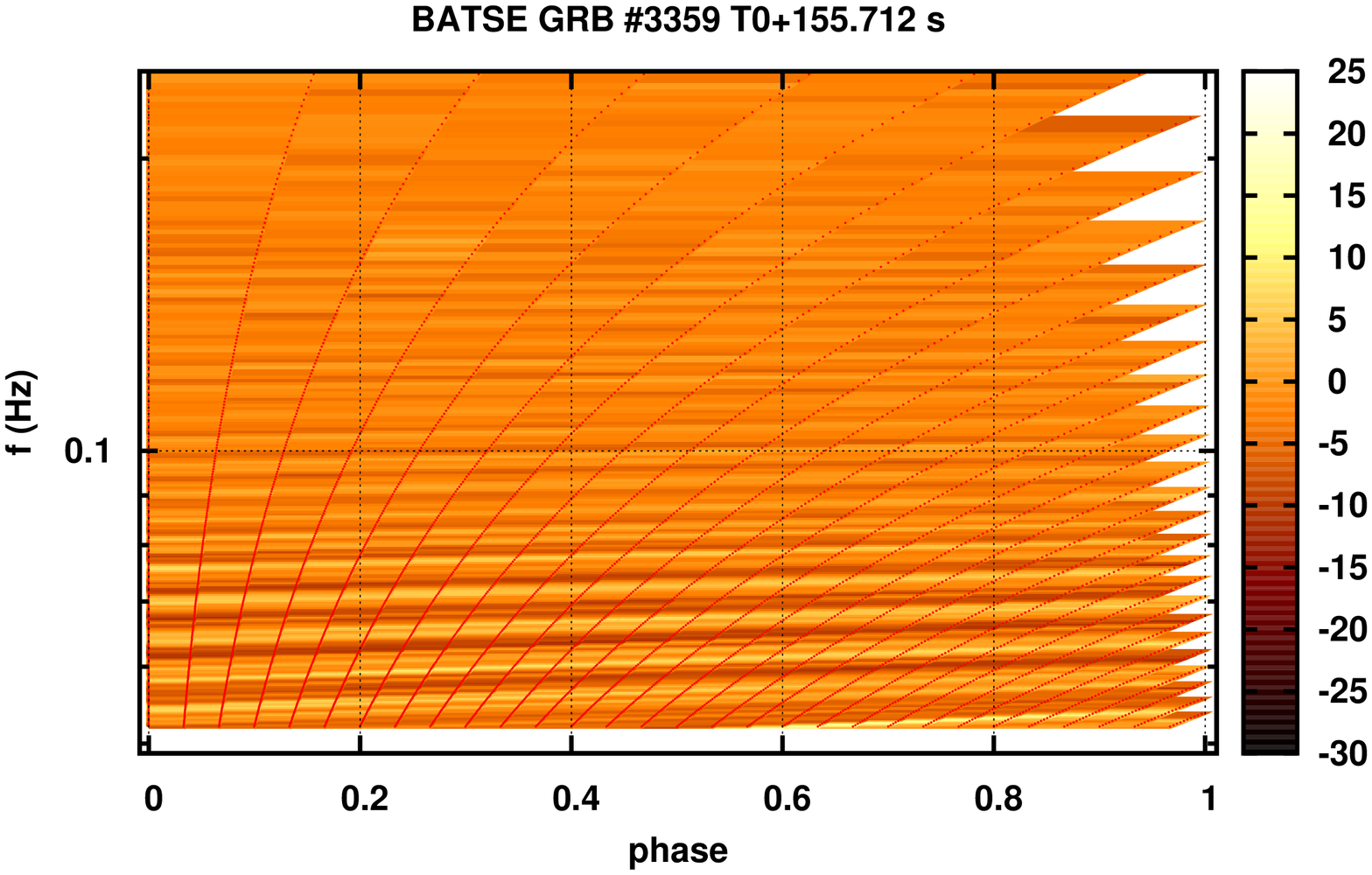} \includegraphics[height=.75\columnwidth, angle=270]{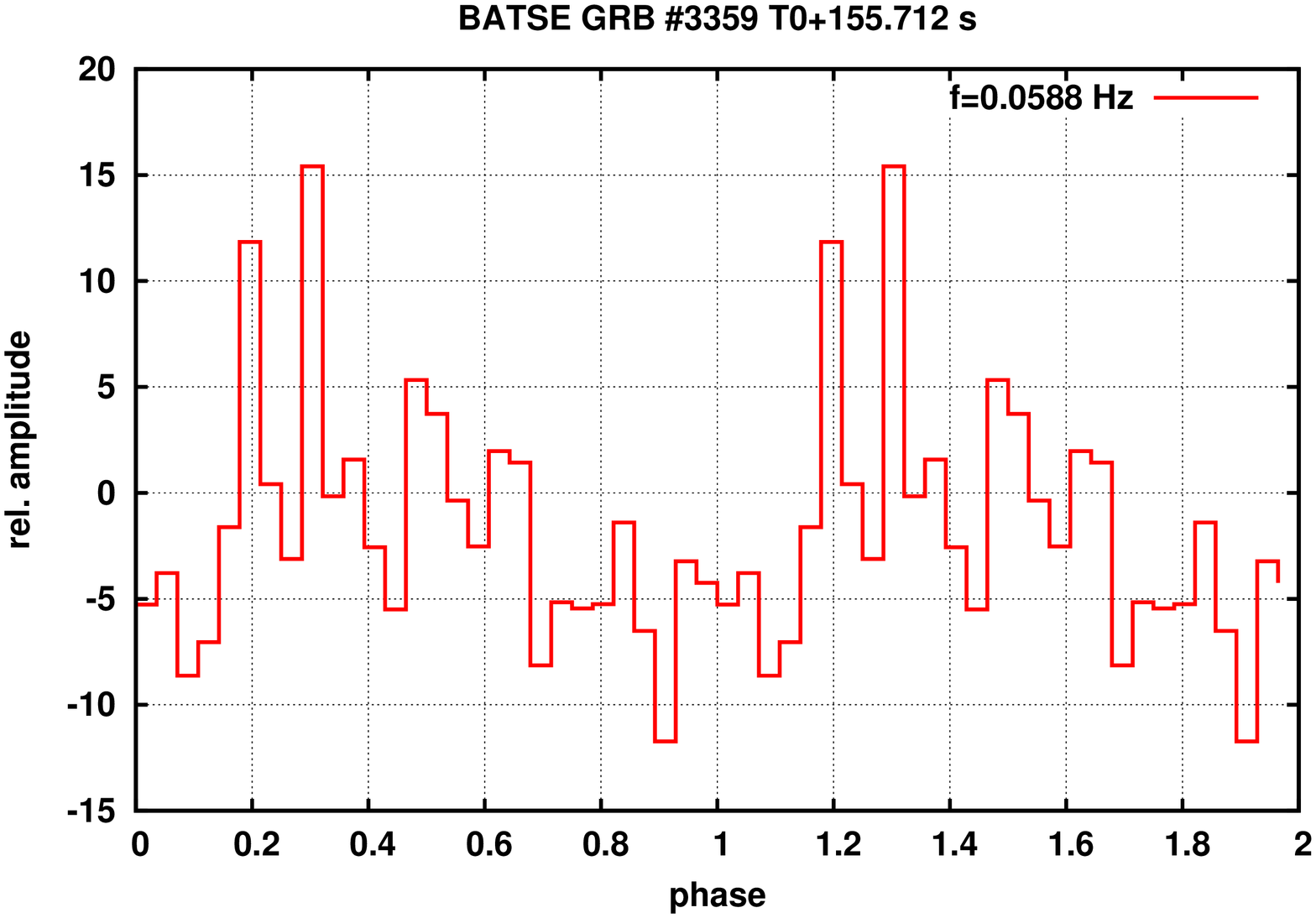}

\includegraphics[height=.75\columnwidth, angle=270]{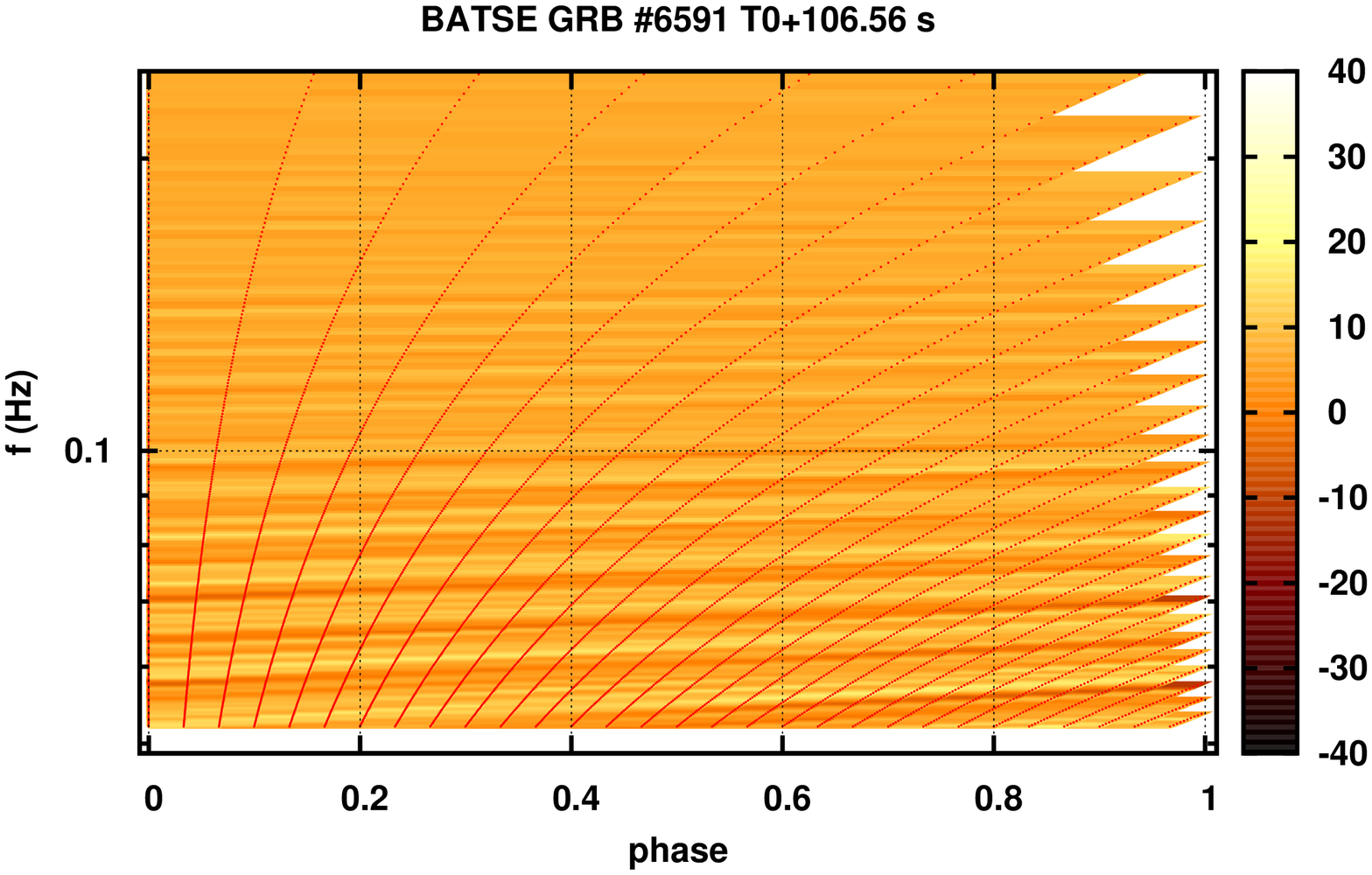} \includegraphics[height=.75\columnwidth, angle=270]{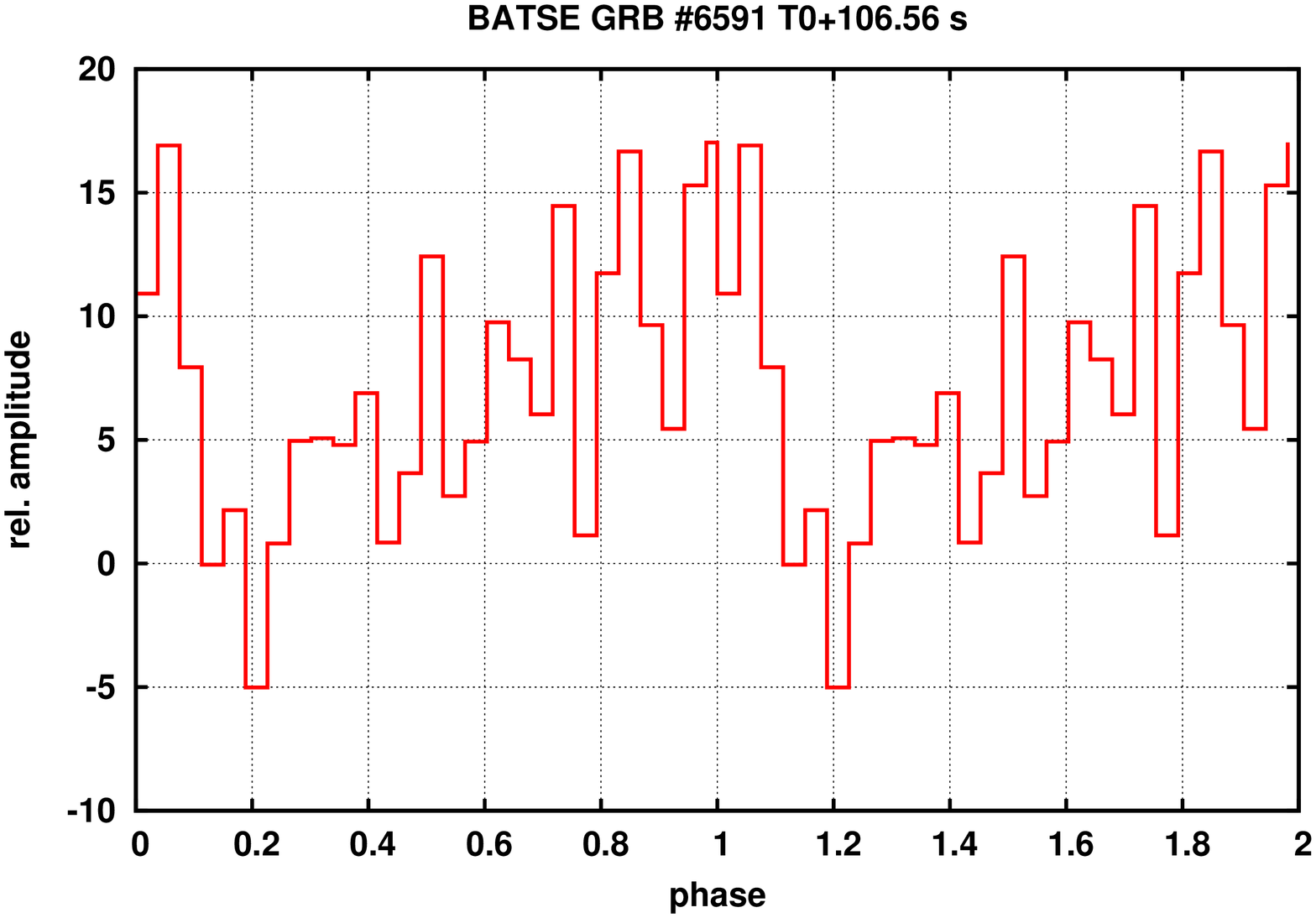}

\bigskip\vspace{0.2cm}
{FIG. 2 (continued): Some folded phase-frequency diagrams and folded pulse shapes of the selected triggers.  }
           \end{figure*}

\section{Sky distribution} 

The sky distribution of the selected 22 triggers is plotted with  on Fig. 3.,
with all the BATSE GRBs' positions. The figure seems to be asymmetric:  to
check the randomness we calculated the dipole components from the selected
subset and from $50000$ cases of a random 22-element sample drawn from the
complete BATSE database: this method fully eliminates the non-uniform
sky-exposure function of the BATSE. We've got only 3 random cases where the
dipole's magnitude was bigger than the 22 triggers' dipole. It gives a
$99.994\%$ significance (approx. $4\sigma$) for the extraordinary dipole
moment.  The direction of the dipole is $l\approx 82^o, b\approx-19^o$.

\begin{figure*}[t]\centering
\includegraphics[height=1.5\columnwidth, angle=270]{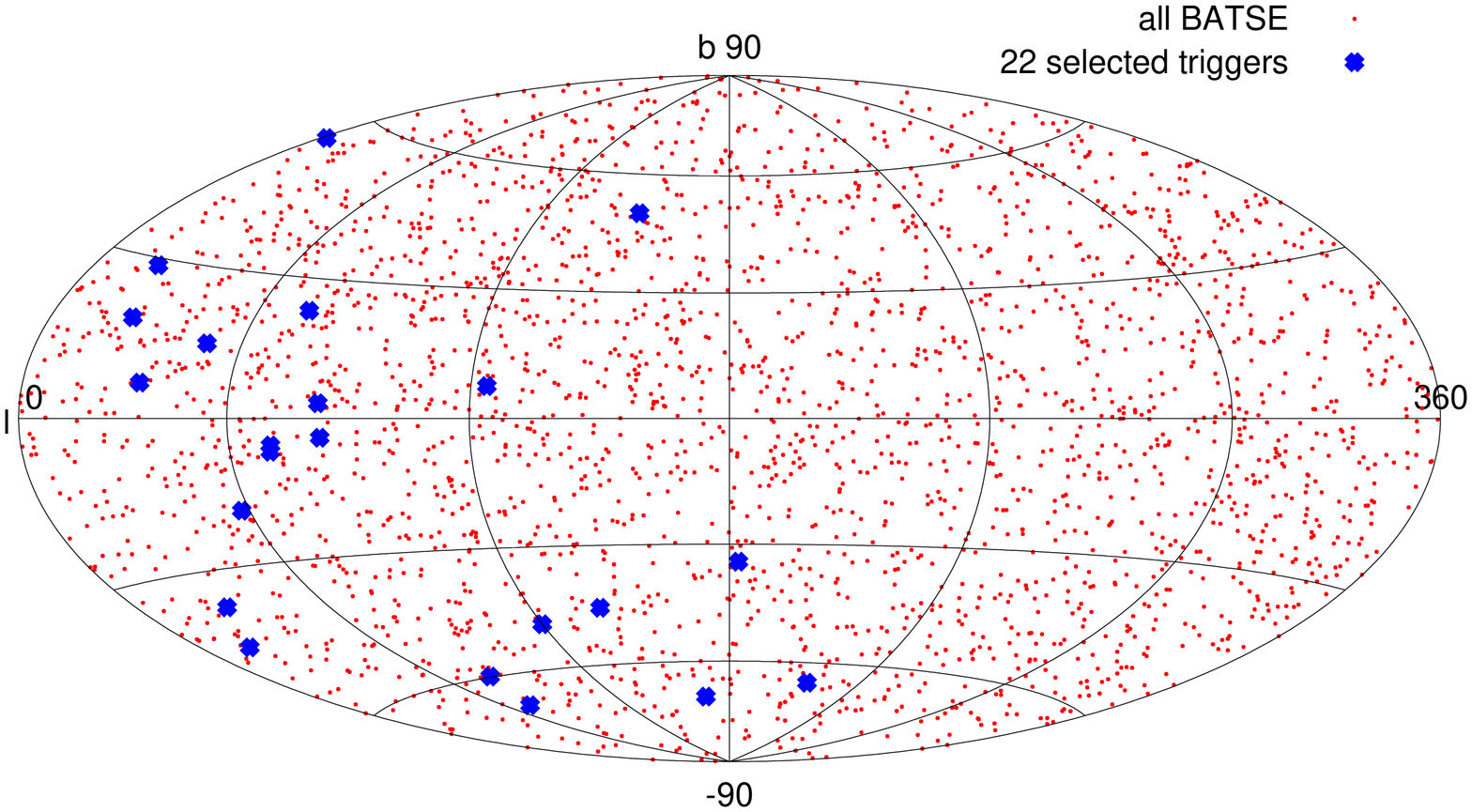}

\bigskip\vspace{0.2cm}
\caption{Sky distribution of the selected triggers. }
\end{figure*}

\section{Discussion}

There are increasing evidence that all the GRBs do not represent a physically
homogeneous group \citep{ho98,muk98,hak00,ho02,bal03,hak03,ho06}.  In the last
years the papers \citep{ba98,me00a,me00b,li01} provided several different tests
probing the sky distribution of GRB groups in BATSE Catalog.  The celestial
distribution of the short BATSE bursts shows anisotropic behavior \citep{me00a,
nor02, vav08}.  The direction of the dipole is not correlated with the
supergalactic plane (not like \cite{nor02}) and interestingly the dipole's
direction is only $\approx 15^o$ away from the the CMB dipole's temperature
minimum.

Our results suggests that some anomalous subgroups could be responsible for the
weak anisotropic signals: for a more detailed answer a deeper study of this
effect is necessary. We plan to provide it in a forthcoming paper.

\section{Acknowledgments}

Thanks are due to  G. Tusn\'ady for valuable discussions.  This research is
supported by Hungarian OTKA grant K077795, by the Bolyai Scholarship (I. H.),
by the GAUK grant No. 46307, and by the Research Program MSM0021620860 of the
Ministry of Education of the Czech Republic (A.M.).

\bigskip 
\bibliographystyle{99}

\end{document}